\documentclass[10pt,twocolumn,twoside]{IEEEtran}

\pdfoutput=1 

\usepackage{etex} 

\ifCLASSINFOpdf
  \usepackage[pdftex]{graphicx}
  \usepackage{epstopdf}
\else
  \usepackage[dvips]{graphicx}
  \graphicspath{{../eps/}}
  \DeclareGraphicsExtensions{.eps}
\fi
\usepackage{cite}

\usepackage[cmex10]{amsmath}
\usepackage{amssymb}
\usepackage{enumerate}
\usepackage{algorithm}
\usepackage{algorithmic}
\usepackage{mdwtab}
\usepackage{amssymb}
\usepackage{eqparbox}
\usepackage{amsfonts}
\usepackage{bm}
\usepackage{booktabs}
\usepackage{bbm}
\usepackage[caption=false,font=footnotesize]{subfig}
\usepackage{url}
\usepackage{acro}
\usepackage{bbding}

\usepackage{tikz}
\usepackage{xcolor}
\usetikzlibrary{decorations.pathreplacing}

\usepackage{balance}

\newcommand{\herm}{^{\mbox{\scriptsize H}}}

\newcommand{\tran}{^{\mbox{\scriptsize T}}}

\newcommand{\vbar}{\raisebox{.17ex}{\rule{.04em}{1.35ex}}}
\newcommand{\vbarind}{\raisebox{.01ex}{\rule{.04em}{1.1ex}}}
\newcommand{\R}{\ifmmode{\rm I}\hspace{-.2em}{\rm R} \else ${\rm I}\hspace{-.2em}{\rm R}$ \fi}
\newcommand{\T}{\ifmmode{\rm I}\hspace{-.2em}{\rm T} \else ${\rm I}\hspace{-.2em}{\rm T}$ \fi}
\newcommand{\N}{\ifmmode{\rm I}\hspace{-.2em}{\rm N} \else \mbox{${\rm I}\hspace{-.2em}{\rm N}$} \fi}
\newcommand{\B}{\ifmmode{\rm I}\hspace{-.2em}{\rm B} \else \mbox{${\rm I}\hspace{-.2em}{\rm B}$} \fi}
\newcommand{\Hil}{\ifmmode{\rm I}\hspace{-.2em}{\rm H} \else \mbox{${\rm I}\hspace{-.2em}{\rm H}$} \fi}
\newcommand{\C}{\ifmmode\hspace{.2em}\vbar\hspace{-.31em}{\rm C} \else \mbox{$\hspace{.2em}\vbar\hspace{-.31em}{\rm C}$} \fi}
\newcommand{\Cind}{\ifmmode\hspace{.2em}\vbarind\hspace{-.25em}{\rm C} \else \mbox{$\hspace{.2em}\vbarind\hspace{-.25em}{\rm C}$} \fi}
\newcommand{\Q}{\ifmmode\hspace{.2em}\vbar\hspace{-.31em}{\rm Q} \else \mbox{$\hspace{.2em}\vbar\hspace{-.31em}{\rm Q}$} \fi}
\newcommand{\Z}{\ifmmode{\rm Z}\hspace{-.28em}{\rm Z} \else ${\rm Z}\hspace{-.28em}{\rm Z}$ \fi}

\renewcommand{\Re}{\mbox{Re}}

\DeclareAcronym{ADMM}{
    short = ADMM,
    long = alternating direction method of multipliers,
    list = Alternating Direction Method of Multipliers,
    class = abbrev
}

\DeclareAcronym{BC}{
    short = BC,
    long = broadcast channel,
    list = Broadcast Channel,
    class = abbrev
}

\DeclareAcronym{BS}{
    short = BS,
    long = base station,
    list = Base Station,
    class = abbrev
}

\DeclareAcronym{BR}{
    short = BR,
    long = best response,
    list = Best Response, 
    class = abbrev
}

\DeclareAcronym{CB}{
    short = CB,
    long = coordinated beamforming,
    list = Coordinated Beamforming,
    class = abbrev
}

\DeclareAcronym{CE}{
    short = CE,
    long = channel estimation,
    list = Channel Estimation,
    class = abbrev
}

\DeclareAcronym{CoMP}{
    short = CoMP,
    long = coordinated multi-point,
    list = Coordinated Multi-Point,
    class = abbrev
}

\DeclareAcronym{CRAN}{
    short = C-RAN,
    long = cloud radio access network,
    list = Cloud Radio Access Network,
    class = abbrev
}

\DeclareAcronym{CSE}{
    short = CSE,
    long = channel specific estimation,
    list = Channel Specific Estimation,
    class = abbrev
}

\DeclareAcronym{CSI}{
    short = CSI,
    long = channel state information,
    list = Channel State Information,
    class = abbrev
}

\DeclareAcronym{CU}{
    short = CU,
    long = central unit,
    list = Central Unit,
    class = abbrev
}

\DeclareAcronym{DE}{
    short = DE,
    long = direct estimation,
    list = Direct Estimation,
    class = abbrev
}

\DeclareAcronym{DE-ADMM}{
    short = DE-ADMM,
    long = direct estimation with alternating direction method of multipliers,
    list = Direct Estimation with Alternating Direction Method of Multipliers,
    class = abbrev
}

\DeclareAcronym{DE-BR}{
    short = DE-BR,
    long = direct estimation with best response,
    list = Direct Estimation with Best Response,
    class = abbrev
}

\DeclareAcronym{DE-SG}{
    short = DE-SG,
    long = direct estimation with stochastic gradient,
    list = Direct Estimation with Stochastic Gradient,
    class = abbrev
}

\DeclareAcronym{DoF}{
    short = DoF,
    long = degrees of freedom,
    list = Degrees of Freedom,
    class = abbrev
}

\DeclareAcronym{DL}{
    short = DL,
    long = downlink,
    list = Downlink,
    class = abbrev
}

\DeclareAcronym{IBC}{
    short = IBC,
    long = interfering broadcast channel,
    list = Interfering Broadcast Channel,
    class = abbrev
}

\DeclareAcronym{i.i.d.}{
    short = i.i.d.,
    long = independent and identically distributed,
    list = Independent and Identically Distributed,
    class = abbrev
}

\DeclareAcronym{JP}{
    short = JP,
    long = joint processing,
    list = Joint Processing,
    class = abbrev
}

\DeclareAcronym{KKT}{
    short = KKT,
    long = Karush-Kuhn-Tucker,
    class = abbrev
}

\DeclareAcronym{LS}{
    short = LS,
    long = least squares,
    list = Least Squares,
    class = abbrev
}

\DeclareAcronym{LTE}{
    short = LTE,
    long = Long Term Evolution,
    class = abbrev
}

\DeclareAcronym{LTE-A}{
    short = LTE-A,
    long = Long Term Evolution Advanced,
    class = abbrev
}

\DeclareAcronym{MIMO}{
    short = MIMO,
    long = multiple-input multiple-output,
    list = Multiple-Input Multiple-Output,
    class = abbrev
}

\DeclareAcronym{MISO}{
    short = MISO,
    long = multiple-input single-output,
    list = Multiple-Input Single-Output,
    class = abbrev
}

\DeclareAcronym{MSE}{
    short = MSE,
    long = mean-squared error,
    list = Mean-Squared Error,
    class = abbrev
}

\DeclareAcronym{MMSE}{
    short = MMSE,
    long = minimum mean-squared error,
    list = Minimum Mean-Squared Error,
    class = abbrev
}

\DeclareAcronym{MU-MIMO}{
    short = MU-MIMO,
    long = multi-user \ac{MIMO},
    list = Multi-User \ac{MIMO},
    class = abbrev
}

\DeclareAcronym{RRH}{
    short = RRH,
    long = remote radio head,
    list = Remote Radio Head,
    class = abbrev
}

\DeclareAcronym{SCA}{
    short = SCA,
    long = successive convex approximation,
    list = Successive Convex Approximation,
    class = abbrev
}

\DeclareAcronym{SG}{
    short = SG,
    long = stochastic gradient,
    list = Stochastic Gradient,
    class = abbrev
}

\DeclareAcronym{SNR}{
    short = SNR,
    long = signal-to-noise ratio,
    list = Signal-to-Noise Ratio,
    class = abbrev
}

\DeclareAcronym{SINR}{
    short = SINR,
    long = signal-to-interference-plus-noise ratio,
    list = Signal-to-Interference-plus-Noise Ratio,
    class = abbrev
}

\DeclareAcronym{SSE}{
    short = SSE,
    long = stream specific estimation,
    list = Stream Specific Estimation,
    class = abbrev
}

\DeclareAcronym{TDD}{
    short = TDD,
    long = time division duplexing,
    list = Time Division Duplexing,
    class = abbrev
}

\DeclareAcronym{UE}{
    short = UE,
    long = user equipment,
    list = User Equipment,
    class = abbrev
}

\DeclareAcronym{UL}{
    short = UL,
    long = uplink,
    list = Uplink,
    class = abbrev
}

\DeclareAcronym{WMMSE}{
    short = WMMSE,
    long = weighted minimum mean-squared error,
    list = Weighted Minimum Mean-Squared Error,
    class = abbrev
}

\DeclareAcronym{WMSEMin}{
    short = WMSEMin,
    long = weighted sum \ac{MSE} minimization,
    list = Weighted sum \ac{MSE} Minimization,
    class = abbrev
}

\DeclareAcronym{WSRMax}{
    short = WSRMax,
    long = weighted sum rate maximization,
    list = Weighted Sum Rate Maximization,
    class = abbrev
}

\newcommand{\ds}{\displaystyle}

\newcommand{\M}[1]{\mathbf{#1}}

\newcommand{\I}{\M{I}}

\begin{document}

\title{Joint Transmission with Limited Backhaul Connectivity}
\author{Jarkko Kaleva, \IEEEmembership{Student Member, IEEE,} 
        Antti T\"olli, \IEEEmembership{Senior Member, IEEE,} 
        Markku Juntti, \IEEEmembership{Senior Member, IEEE,} 
        Randall Berry, \IEEEmembership{Fellow, IEEE} and 
        Michael Honig \IEEEmembership{Fellow, IEEE} 
    
\thanks{Jarkko Kaleva, Antti T\"olli and Markku Juntti are with Centre for
    Wireless Communications, University of Oulu P.O. Box 4500, 90014 University
    of Oulu, Finland. Email: jarkko.kaleva@ee.oulu.fi; antti.tolli@ee.oulu.fi;
    markku.juntti@ee.oulu.fi. Randall Berry and Michael Honig are with
    Northwestern University, Dept.\ of Elect.  Engineering and Comp. Science,
    Northwestern University, Evanston, IL, 60208, U.S.
    
    Parts of this work has been published in IEEE International Conference on
    Acoustics, Speech, and Signal Processing, Florence, Italy, 2014  and IEEE 
    Global Communications Conference, San Diego, U.S,  2015.}
}

\maketitle

\begin{abstract}
    Downlink beamforming techniques with low signaling overhead are proposed for
    joint processing coordinated (JP) multi-point transmission. The objective is to
    maximize the weighted sum rate within joint transmission clusters. As the
    considered weighted sum rate maximization is a non-convex problem,
    successive convex approximation techniques, based on weighted mean-squared
    error minimization, are applied to devise algorithms with tractable
    computational complexity. Decentralized algorithms are proposed to
    enable JP even with limited backhaul connectivity.
    These algorithms rely provide a variety of
    alternatives for signaling overhead, computational complexity and
    convergence behavior. Time division duplexing is exploited to design
    transceiver training techniques for two scenarios: stream specific
    estimation and direct estimation. In the stream specific estimation, the
    base station and user equipment estimate all of the stream specific precoded
    pilots individually and construct the transmit/receive covariance matrices
    based on these pilot estimates. With the direct estimation, only the
    intended transmission is separately estimated and the covariance matrices
    constructed directly from the aggregate system-wide pilots. The proposed
    training schemes incorporate bi-directional beamformer signaling to improve
    the convergence behavior. This scheme exploits the time division duplexing
    frame structure and is shown to improve the training latency of the
    iterative transceiver design.  The impact of feedback/backhaul signaling
    quantization is considered, in order to further reduce the signaling
    overhead. Also, user admission is being considered for time-correlated
    channels.  The enhanced transceiver convergence rate enables periodic
    beamformer reinitialization, which greatly improves the achieved system
    performance in dense networks.
\end{abstract}


\IEEEpeerreviewmaketitle


\section{Introduction}

\IEEEPARstart{C}{ooperative} transmission schemes and spatial domain
interference managements are the foundation of the modern cellular and
heterogeneous wireless systems. The ever increasing need for spectral
efficiency, imposes demand for effective interference management and
transmission coordination.  The current wireless standards already support
efficient single cell \ac{MIMO} beamforming, which allows smart beamformer
design to efficiently take advantage of the multi-user diversity in the spatial
domain~\cite{Dahlman-Parkvall-Skold-2011}. Also, the basic operation of virtual
\ac{MIMO} has been preliminarily covered by the \ac{LTE-A} standards. This also
includes support for fully centralized \ac{JP} \ac{CoMP} transmission schemes,
which from the theoretical perspective does not differ much from single cell
\ac{MIMO} beamformer processing. Still, the practical limitations in \ac{BS}
backhaul connectivity are preventing effective implementation of more advanced
\ac{JP} \ac{CoMP} schemes. Multi-cell beam coordination is still in somewhat
elementary stages when considering the current \ac{LTE-A}
standardization~\cite{Dahlman-Parkvall-Skold-2011}. On the other hand, a lot of
research effort has been invested in \ac{CB} for multi-cell systems. Much of
this research has been focusing on decentralized coordination strategies with
low backhaul signaling overhead in mind~\cite{Shi-Razaviyayn-Luo-He-11,
Komulainen-Tolli-Juntti-13, Tolli-Pennanen-Komulainen-TWC-11,
Bogale-Vandendorpe-11, Pennanen-Tolli-Latva-Aho-11}. These techniques are still
limited in the \ac{DoF} sense and fall short in exploiting the available
opportunities of the increasingly dense cell networks.  Most of the \ac{CB}
research effort is focused on mitigating and managing the inter-cell
interference, as it is the foremost barrier preventing efficient spectrum
utilization especially, in dense heterogeneous networks. To this end, \ac{JP}
\ac{CoMP} transmission techniques have been considered, where the neighboring
\acp{BS} cooperate on different levels, in order to increase the available
\ac{DoF} and alleviate the detrimental interference conditions in multi-cell
systems~\cite{Gesbert-Hanly-Huang-Shamai-Simeone-Wei-10, Zhou-Gong-Niu-WC-11,
Lee-Seo-Clerckx-Hardouin-Maazarese-Nagata-Sauana-12}.

\ac{JP} \ac{CoMP}, in general, requires high bandwidth and low latency \ac{BS}
inter-connectivity, which is hard to realize in conventional multi-cell systems.
This has led to development of network architectures that make virtual \ac{MIMO}
operation realizable and offer the required centralized coordination. The most
popular architectures in today's \ac{JP} research are the
\acp{CRAN}~\cite{I-Rowell-Han-Zu-Li_Pan-14}. In \ac{CRAN} the \acp{BS} are
connected to a \ac{CU} over high capacity and low latency backhaul links. This
allows the \ac{CU} to perform centralized processing and the \acp{BS} act merely
as \acp{RRH}. The \ac{CU} can use \ac{JP} to utilize simultaneously multiple
\acp{RRH} for beamforming. Although the backhaul limitations of \ac{CoMP}
transmission systems have been addressed in various publications, most of the
\ac{JP} \ac{CoMP} research allows full \ac{CSI} exchange and centralized
processing which greatly simplifies the beamformer
design~\cite{Zhang-Andrews-10, Zhou-Gong-Niu-WC-11,
    Tolli-Pennanen-Komulainen-TWC-11, Han-Yang-Wang-Zhy-Lei-13,
Kim-Sun-Paulraj-13}. While the global \ac{CSI} exchanged is common assumption
for \ac{JP} designs, in many cases, it may not be feasible in practice. The
latency and mobility requirements often prevent accurate \ac{CSI} exchange even
in modest scale~\cite{Lee-Seo-Clerckx-Hardouin-Maazarese-Nagata-Sauana-12}. 

This paper focuses on \ac{JP} \ac{CoMP} with \ac{WSRMax} and limited backhaul
capacity, i.e., without full \ac{CSI} exchange. The backhaul limitations prevent
the \acp{BS} from sharing the global \ac{CSI} within the cooperating clustering.
Thus, the limited signaling possibilities require the \acp{BS} to perform
partially independent beamformer design even within the cooperating \ac{JP}
clustering. This has motivated us to design decentralized transceiver processing
with limited signaling overhead. Still, we assume that the transmitted data can
be shared among the serving \acp{BS}. The data can be queued and prioritized by
a central processing node, which then distributes it to the serving \acp{BS}.
This makes the data less sensitive to the latency in the system. 

Along with low signaling overhead, we also consider different transceiver
training techniques depending on the pilot planning and contamination when using
\ac{TDD}. Basically, we divide the problem into two scenarios: \ac{SSE} and
\ac{DE}. In \ac{SSE}, we assume that all pilot sequences are orthogonal and
orthogonal. With \ac{DE}, we allow non-orthogonal noisy pilots, which is
sensible in more crowded environments.

\subsection{Prior Work}

\ac{CB} has extensively studied with respect to decentralized inter-cell
interference coordination. In \ac{CB}, the interfering \acp{BS} cooperatively
coordinate the beamformers and schedule their \acp{UE}, is such a way that the
inter-cell interference conditions are not detrimental for the neighboring
cells. \Ac{CB} can be efficiently performed with limited signaling overhead as
shown, e.g., in~\cite{Tolli-Pennanen-Komulainen-TWC-11,
Shi-Razaviyayn-Luo-He-11, Komulainen-Tolli-Juntti-13}. \ac{WSRMax} with
\ac{CB} have been studied, e.g., in~\cite{Shi-Schubert-Boche-08,
Codreanu-Tolli-Juntti-Latva-aho-trsp-07,
Christensen-Agarwal-Carvalho-Cioffi-TWC-08}. Over the last ten years, many
practical \ac{CB} schemes have been proposed with reasonable signaling overhead
and computational complexity.  The most popular approach is via the, so called,
\ac{WMMSE} design, where the problem is equivalently presented as logarithmic
\ac{MSE} minimization problem. This problem is still non-convex and \ac{SCA} is
applied on the non-convex objective. This results in iteratively weighted
\ac{MSE} minimization, which can be efficiently solved. The \ac{WMMSE} method
was first proposed in~\cite{Christensen-Agarwal-Carvalho-Cioffi-TWC-08}.
In~\cite{Shi-Razaviyayn-Luo-He-11}, \ac{WMMSE} was shown to have naturally
decentralized processing structure for cellular \ac{TDD} \ac{MIMO} systems.
Efficient signaling techniques for \ac{WMMSE} were proposed
in~\cite{Komulainen-Tolli-Juntti-13}. Similar approach to \ac{WSRMax} has also
been considered in~\cite{Bogale-Vandendorpe-11,
Scutari-Facchinei-Song-Palomar-Pang-14, Kaleva-Tolli-Juntti-TSP16}.  The
methods proposed in~\cite{Christensen-Agarwal-Carvalho-Cioffi-TWC-08,
Shi-Razaviyayn-Luo-He-11} are readily applicable to jointly coherent
centralized \ac{CoMP} beam coordination, as the problem closely resembles a
conventional single-cell coordinated \ac{WSRMax} problem. Since \ac{JP}
inherently couples the beamformer processing between the cooperating
\acp{BS}, the decentralized \ac{CB} methods cannot be used as is.
Developing the decentralized \ac{CoMP} designs is one of the focus points of
this paper.

In \ac{JP}, the backhaul information can be handled in one of two ways: 1. In
data sharing, the \ac{CU} exchanges the user specific messages with
the ̧cooperating \acp{BS} explicitly and the joint beamformers are informed
separately~\cite{Hong-Sun-Baligh-Luo-2013}. 2. Using compression, the messages
can be precoded beforehand at the \ac{CU} and only the compressed version of the
analog beamformer is informed to the
\acp{BS}~\cite{Park-Simeone-Sahin-Shamai-13, Dai-Yu-16}.  As of now, the data
sharing strategy has been more popular approach, mostly due to the lower
complexity and easier modeling of the explicit backhaul constraints. Sparsity
imposing joint beamformer designs are the most common approach for backhaul
limited \ac{CoMP} design. The sparse joint beamforming has been considered,
e.g., in~\cite{Hong-Sun-Baligh-Luo-2013, Zhuang-Lau-2014, Lia-Hong-Liu-Luo-2014,
Kaleva-Bande-Tolli-Juntti-Veeravalli-Spawc16}. These designs try to limit the
sizes of the \ac{JP} clusters and, thus, implicitly reduce the backhaul
overhead. The compression approach has recently gained more
popularity~\cite{Park-Simeone-Sahin-Shamai-13,  Park-Simeone-Sahin-Shamai-14,
Park-Siemone-Shain-Shamai-14_MAG, Dai-Yu-16} Different aspects and benefits of
backhaul compression have been studied, for example,
in~\cite{Park-Simeone-Sahin-Shamai-13,  Park-Simeone-Sahin-Shamai-14,
Park-Siemone-Shain-Shamai-14_MAG,Dai-Yu-16}.  The data sharing and compression
strategies for energy efficient communication and backhaul power consumption
were compared in~\cite{Dai-Yu-16}.

Decentralized interference management has been a major research interest in the
cellular beamformer design. The backhaul delay and capacity limitations have
motivated \ac{CB} research to find solutions beyond the centralized processing
concepts. Most of this research has focused on decomposition techniques of
convex beamformer optimization problems~\cite{Tolli-Pennanen-Komulainen-TWC-11,
Shen-Chang-Wang-Qiu-Chi-12}. For \ac{CB}, the \ac{WSRMax} has been shown to have
decoupled structure when \ac{SCA} is
applied~\cite{Christensen-Agarwal-Carvalho-Cioffi-TWC-08,
Shi-Razaviyayn-Luo-He-11}. More generalized frameworks have also been proposed.
For example, a general \ac{BR} framework, which allows straightforward parallel
processing for varying performance objectives, was proposed
in~\cite{Scutari-Facchinei-Song-Palomar-Pang-14}.


Pilot non-orthogonality and contamination has been widely studied, albeit, not
in the context of \ac{JP} \ac{CoMP}. More generally, the impact of imperfect
\ac{CSI} has been a popular topic in the literature. The impact of partial or
imperfect \ac{CSI} feedback to \ac{CB} and \ac{JP} has been studied in various
publications, e.g.,~\cite{Zhou-Gong-Niu-WC-11,Kim-Shin-Sohn-Lee-12,
Jaramillo-Ramirez-Kountouris-Hardouin-15}. Partially available \ac{CSI} imposes
a different problem to the one considered herein. With imperfect \ac{CSI}, the
problem is more sensitive to the management of the \ac{CSI} uncertainty. The
pilot contamination in \ac{TDD} based transceiver training for coordinated
beamforming has been considered, e.g.,
in~\cite{Jose-Ashikmin-Marzetta-Vishwanath-WC-11, Shi-Berry-Honig-TSP-14,
Xu-Guo-Honig-TSP-15}. In~\cite{Shi-Berry-Honig-TSP-14, Xu-Guo-Honig-TSP-15},
direct \ac{LS} beamformer estimation from the contaminated \ac{UL}/\ac{DL}
pilots was shown to provide good performance as opposed to trying to estimate
the individual channels.

\subsection{Contributions}

We provide \ac{WSRMax} design \ac{JP} \ac{CoMP} with emphasis on systems with
moderately fast fading channel conditions We assume data sharing, where the
\ac{CU} provides the data for cooperating \acp{BS}. The transceiver processing
is done with minimal \ac{CU} involvement by novel decentralized \ac{CoMP}
beamformer designs.  Our focus is on practically realizable signaling schemes
and efficient user admission. We extend the decentralized sum-\ac{MSE}
minimizing \ac{JP} scheme proposed
in~\cite{Kaleva-Berry-Honig-Tolli-Juntti-ICASSP-14} to perform \ac{WMMSE} with
multi-antenna transmitters and receivers. We employ \ac{BR}, \ac{ADMM} and
\ac{SG} schemes to provide decentralized algorithms with different performance
properties and signaling overhead. We also consider pilot contamination with
\ac{DE} methods and provide efficient decentralized processing via \ac{SG} for
the case with imperfect channel estimation as well. We utilize a bi-directional
signaling scheme with similar frame structure as
in~\cite{Shi-Berry-Honig-TSP-14}. This allows fast signaling iterations by
incorporating the training sequence into the transmitted frame structure.
Furthermore, we propose a novel periodic beamformer reinitialization scheme,
which is shown to significantly improve the user admission performance in time
correlated fading environment.  The performance of the proposed method is
evaluated in a cellular multi-user network with a time correlated channel model.

Contributions of this paper are summarized as follows: 
\begin{itemize} 
    \item \ac{BR}, \ac{ADMM} and \ac{SG} based decentralized beamforming are
        proposed for stream specific \ac{SSE}. 
    \item Beamformer \ac{DE} with \ac{SG} based decentralized beamforming
        procedures are considered. 
    \item User admission, bi-directional training and feedback quantization
        techniques are considered for time-correlated \ac{CoMP} processing. 
    \item Performance of the proposed methods are studied with numerical
        examples in time correlated channels. 
\end{itemize}

\subsection{Organization and Notation}
The rest of the paper is organized as follows. The system model is given in
Section~\ref{sec:sysmodel}.  In Section~\ref{sec:problem}, the considered
\ac{WSRMax} problem is described along with the \ac{WMMSE} \ac{SCA} design.  The
beamformer designs with \ac{SSE} are given in Section~\ref{sec:sse}. \ac{DE} is
considered in Section~\ref{sec:direct_estimation}. Bi-directional training and
feedback quantization are discussed in Section~\ref{sec:training}. Finally, the
numerical examples and concluding remarks are given in Sections~\ref{sec:simres}
and~\ref{sec:conclusions}, respectively.

{\it Notation:\/} Matrices and vectors are presented by boldface upper and lower
case letters, respectively.  Transpose of matrix $\mathbf{A}$ is denoted as
$\mathbf{A}\tran$ and, similarly, conjugate transpose is denoted as
$\mathbf{A}\herm$. Conventional matrix inversion is written as
$\mathbf{A}^{-1}$, whereas $\mathbf{A}^{\dagger}$ presents the pseudoinverse of
matrix $\mathbf{A}$. Mapping of negative scalars to zero is written as
${(\cdot)}^{+} = \max(0, \cdot)$. Cardinality of a discrete set $\mathcal{A}$ is
given by $|\mathcal{A}|$. Expected value of a random variable is denoted by
$\mathbb{E}[\cdot]$.


\section{System Model}
\label{sec:sysmodel}

\begin{figure}[ht]
    \centering
    \includegraphics[width=1.0\columnwidth]{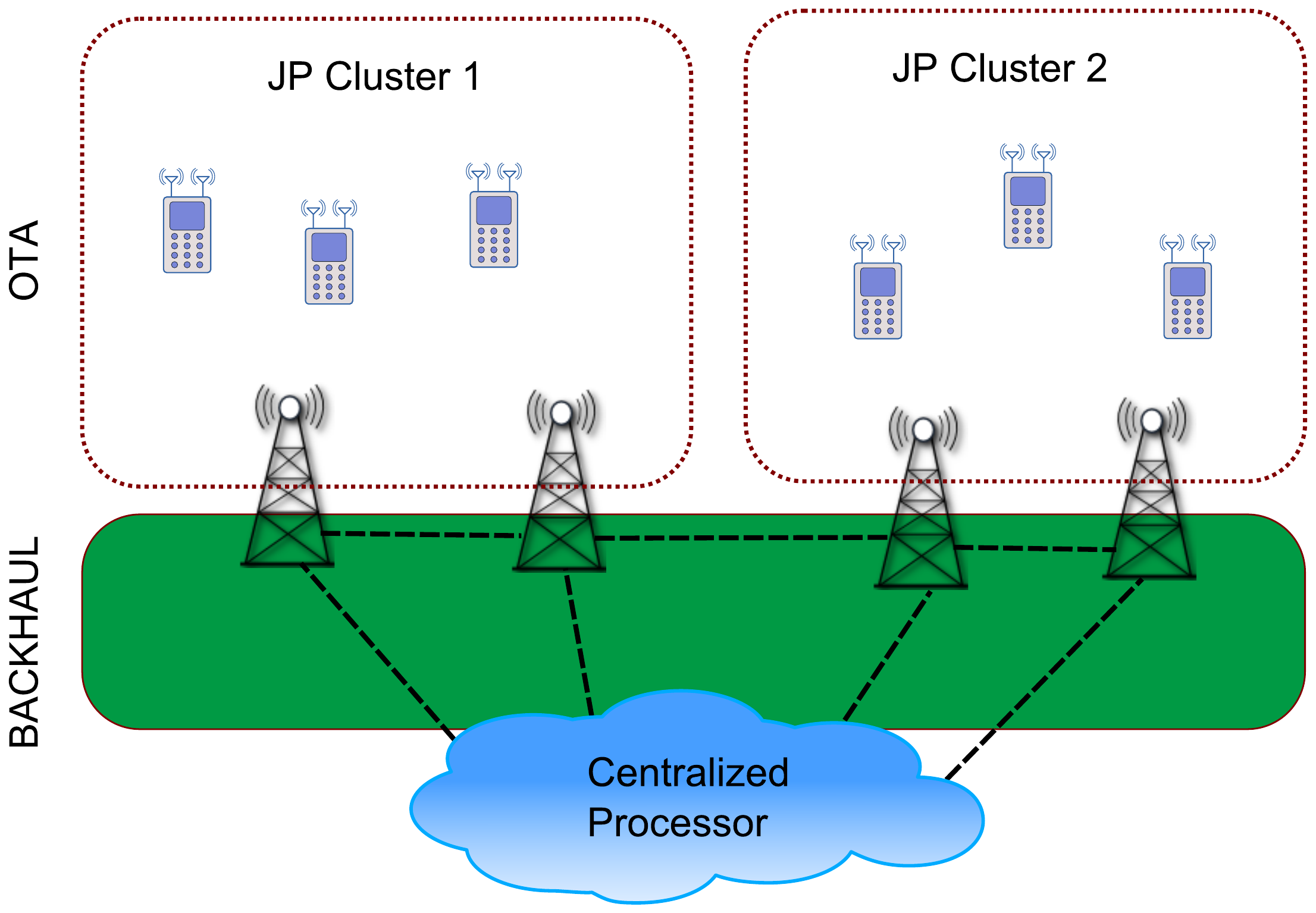}
        \caption{Simplified system model with $B = 4$ cooperating \acp{BS} and
            $K = 6$ \acp{UE} that are split into two \ac{JP} clusters.}
\label{fig:sysmodel}
\end{figure}

We consider a multi-cell system with $B$ \acp{BS} each equipped with
$N_\text{T}$ transmit antennas. There are, in total, $K$ \acp{UE} each equipped
with $N_\text{R}$ receive antennas. Each \ac{UE} $k = 1,\ldots,K$ is coherently
served by $|\mathcal{B}_k|$ \acp{BS}, where set $\mathcal{B}_k$  defines the
joint processing cluster (coherent serving \ac{BS} indices) for \ac{UE} $k$.
Similarly, the set of \ac{UE} indices served by \ac{BS} $b = 1,\ldots,B$ is
denoted by $\mathcal{C}_b = \{k | b \in \mathcal{B}_k, k = 1,\ldots,K\}$.  The
set of all \ac{UE} indices is given by $\mathcal{K} = \{1,\ldots,K\}$. The
maximum number of spatial data streams allocated to \ac{UE} $k = 1,\ldots,K$ is
denoted by $L_k \leq \min\left(|\mathcal{B}_k| N_\text{T}, N_\text{R}\right)$.
To simplify the notation in various places, we use the following set
abbreviations: $(k, l) \triangleq \{(k, l) | k \in \mathcal{K}, l =
1,\ldots,L_k\}$ and $(b, k, l) \triangleq \{(b, k, l) | k \in \mathcal{K}, l =
1,\ldots,L_k, b \in \mathcal{B}_k\}$. The considered system model is illustrated
in Fig.~\ref{fig:sysmodel}.

The downlink transmission within the \ac{JP} set is considered to be symbol
synchronous in the sense that each transmitted symbol from $\mathcal{B}_k, k =
1,\ldots,K$ is coherently combined at all \acp{UE}.  Only the local \ac{CSI}
knowledge is assumed, that is, each \ac{BS} $b = 1,\ldots,B$ is only aware of
the channel matrix $\M{H}_{b,k} \in \mathbb{C}^{N_\text{R} \times N_\text{T}}\
\forall\ k = 1,\ldots,K$, while the data sharing is assumed within each serving
set of \acp{BS} $\mathcal{B}_k$.  Furthermore, we assume \ac{TDD}, which imposes
strong correlation between the \ac{UL} and \ac{DL} channels.

The received signal at \ac{UE} $k = 1,\ldots,K$ is given as
\begin{equation}
    \M{y}_k = \sum_{i = 1}^K\sum_{b\in\mathcal{B}_i}\sum_{j = 1}^{L_i}
        \M{H}_{b,k}\M{m}_{b,i,j}d_{i,j} + \M{n}_k
    \text{,}
\end{equation}
where $\M{m}_{b,i,j} \in \mathbb{C}^{N_\text{T}}$ is the beamformer vector for
the $j^{th}$ spatial data stream of \ac{UE} $i$ from \ac{BS} $b$ and $\M{n}_k
\sim \mathcal{CN}(0, \sigma_k^2\M{I})$ denotes the receiver noise.  The complex
data symbols $d_{k,l}, k = 1,\ldots,K, l = 1,\ldots,L_k$ are assumed to be
\ac{i.i.d.} with $\mathbb{E}\{|d_{k,l}|^2\} = 1$.

The estimated symbol at \ac{UE} $k = 1,\ldots,K$ over stream $l$, after the
applying receive beamformer $\M{u}_{k,l} \in \mathbb{C}^{N_\text{R}}$, is given
as $\hat{d}_{k,l} = \M{u}_{k,l}\herm \M{y}_k$. The resulting \ac{SINR} is 
\begin{equation}
    \begin{array}{rl}
    \label{eq:sinr}
    \ds \Gamma_{k,l} =
        \frac{\ds |\sum_{b\in\mathcal{B}_k}
                    \M{u}_{k,l}\herm\M{H}_{b,k}\M{m}_{b,k,l}|^2}
             {\ds \sum_{i = 1}^K\sum_{\substack{j = 1,\\ (i,j)\neq(k,l)}}^{L_i}
                |\sum_{b\in\mathcal{B}_i}
                    \M{u}_{k,l}\herm\M{H}_{b,k}\M{m}_{b,i,j}|^2 + 
                \|\M{u}_{k,l}\|^2\sigma^2_k}
    \end{array}
    \text{,}
\end{equation}
and the corresponding \ac{MSE} is 
\begin{equation}
    \begin{array}{rl}
    \label{eq:mse}
    \ds \epsilon_{k,l} \triangleq & 
        \ds \mathbb{E}[|d_{k,l} - \hat{d}_{k,l}|^2 ] \\
        = & \ds
        |\sum_{b\in\mathcal{B}_k} 
            \M{u}_{k,l}\herm\M{H}_{b,k} \M{m}_{b,k,l} - 1|^2 + 
        \|\M{u}_{k,l}\|^2\sigma_k^2 +
    \\ 
    & \ds
        \sum_{i = 1}^K\sum_{\substack{j = 1,\\ (i,j) \neq (k,l)}}^{L_i}
            |\sum_{b\in\mathcal{B}_i} 
                \M{u}_{k,l}\herm\M{H}_{b,k} \M{m}_{b,i,j}|^2 
        \text{.}
    \end{array}
\end{equation}
Note that~\eqref{eq:mse} is a convex function in terms of the transmit
beamformers $\M{m}_{b,k,l}\ \forall\ (b,k,l)$ for fixed receivers $\M{u}_{k,l}\
\forall\ (k,l)$.


\section{Problem Formulation \& Centralized Solution}
\label{sec:problem}

We consider \ac{WSRMax} subject to \ac{BS} specific sum transmit power
constraints. The general problem can be given as
\begin{equation}
    \label{prob:wsrmax}
    \begin{array}{rl} 
        \underset{\substack{\mathbf{u}_{k,l}, \mathbf{m}_{b,k,l}}}{\max.} &  
        \ds 
            \sum_{k = 1}^K\sum_{l = 1}^{L_k} 
                \mu_k \log_2\left(1 + \Gamma_{k,l}\right)
                \\
        \mathrm{s.\ t.} 
            & \ds
                \sum_{k \in \mathcal{C}_b}^K\sum_{l = 1}^{L_k} 
                    \|\mathbf{m}_{b,k,l}\|^2 \leq P_b,\ b = 1,\ldots,B
            \text{,}
    \end{array}
\end{equation}
where $\mu_k, k = 1,\ldots,K$ are the user priority weights.  The problem is
non-convex and known to be NP-hard~\cite{Luo-Zhang-08}. The optimal, i.e., rate
maximizing, receive beamformers for~\eqref{prob:wsrmax} are the \ac{MMSE}
receive beamformers
\begin{equation}
\label{eq:lmmse}
        \mathbf{u}_{k,l} = 
            \mathbf{K}_k^{-1}
                \left(
                    \sum_{b \in \mathcal{B}_k}
                        \mathbf{H}_{b,k} \mathbf{m}_{b,k,l}
                \right)
        \text{,}
\end{equation}
where 
\begin{math}
    \mathbf{K}_{k} =
        \sum_{i = 1}^K\sum_{j = 1}^{L_i}\sum_{b\in\mathcal{B}_i} 
            \mathbf{H}_{b,k} \mathbf{m}_{b,i,j}
            \mathbf{m}_{b,i,j}\herm \mathbf{H}_{b,k}\herm + 
            \mathbf{I}\sigma_k^2
            \text{.}
\end{math}

It is well-known that, when the \ac{MMSE} receive beamformers are applied, there
is an inverse relation between the \ac{SINR} and the corresponding
\ac{MSE}~\cite{Shi-Razaviyayn-Luo-He-11}
\begin{equation}
    \label{eq:mse_sinr}
    \epsilon_{k,l}^{-1} = 1+ \Gamma_{k,l}
    \text{.}
\end{equation}

Now, applying~\eqref{eq:mse_sinr} to~\eqref{prob:wsrmax} we can formulate the
weighted sum rate maximization problem as\footnote{$\log_2(1 + \Gamma_{k,l}) =
\log_2(\epsilon_{k,l}^{-1}) = -\log_2(\epsilon_{k,l})$.}
\begin{equation}
\label{prob:mse_min}
    \begin{array}{rl} 
        \underset{\substack{\mathbf{u}_{k,l}, \mathbf{m}_{b,k,l}}}{\min.} &  
        \ds 
            \sum_{k = 1}^K\sum_{l = 1}^{L_k} 
                \mu_k\log_2\left(\epsilon_{k,l}\right)
                \\
        \mathrm{s.\ t.} 
            & \ds
                \sum_{k \in \mathcal{C}_b}^K\sum_{l = 1}^{L_k} 
                    \|\mathbf{m}_{b,k,l}\|^2 \leq P_b,\ b = 1,\ldots,B
            \text{.}
    \end{array}
\end{equation}

Since~\eqref{prob:mse_min} is not jointly convex for the transmit and receive
beamformers, we consider an alternating design, where the transmit and receive
beamformers are solved in separate instances. This separation is convenient for
\ac{TDD} processing as the \ac{DL} and \ac{UL} transmissions are temporally
separated.  With fixed transmit beamformers $\M{m}_{b,k,l}\ \forall\ (b,k,l)$,
the optimal receive beamformer can be obtained
from~\eqref{eq:lmmse}~\cite{Christensen-Agarwal-Carvalho-Cioffi-TWC-08}. This
can be easily verified from the first-order \ac{KKT} conditions.
As~\eqref{prob:mse_min} is still non-convex, even for fixed receive beamformers,
we need to apply an iterative convex approximation algorithm to generate the
transmit beamformers. We employ
\ac{WMMSE}~\cite{Christensen-Agarwal-Carvalho-Cioffi-TWC-08,
Shi-Razaviyayn-Luo-He-11} to formulate transmit beamformer design algorithm with
tractable complexity and convex steps.

The main idea in the \ac{WMMSE} is to perform successive first-order
approximation of the non-convex objective in~\eqref{prob:mse_min}. The objective
is separable in terms of $\epsilon_{k,l}\ \forall\ (k,l)$. Thus, we can
approximate each term individually. In iteration $n$, the first-order
approximation in terms of $\epsilon_{k,l}$, at point $\epsilon_{k,l}^{(n)}$, can
be given as
\begin{equation}
\label{eq:mse_approx}
    \log_2 \left(\epsilon_{k,l}\right) \approx 
        \frac{1}{\log(2)\epsilon_{k,l}^{(n)}}
        \left(\epsilon_{k,l} - \epsilon_{k,l}^{(n)}\right) + 
        \log_2\left(\epsilon_{k,l}^{(n)}\right) 
\text{.}
\end{equation}
When optimizing over the approximated objective, the constant terms have no
impact on the solution and can, thus, be neglected. Now, the approximated
transmit beamformer design subproblem can be given as a \ac{WMMSE} problem
\begin{equation}
\label{prob:wmmse_min}
    \begin{array}{rl} 
        \underset{\substack{\M{m}_{b,k,l}}}{\min.} &  
        \ds 
            \sum_{k = 1}^K\sum_{l = 1}^{L_k} w^{(n)}_{k,l}\epsilon_{k,l}
    \\
    \mathrm{s.\ t.} 
    & \ds
        \sum_{k \in \mathcal{C}_b}^K\sum_{l = 1}^{L_k} 
            \|\M{m}_{b,k,l}\|^2 \leq P_b,\ b = 1,\ldots,B
    \text{,}
\end{array}
\end{equation}
where 
\begin{equation}
\label{eq:mse_weights}
    w^{(n)}_{k,l} = \frac{\mu_k}{\log(2)\epsilon_{k,l}^{(n)}}\ \forall\ (k,l)
    \text{.}
\end{equation}
Since the \ac{MSE} terms~\eqref{eq:mse} are convex for fixed receive
beamformers,~\eqref{prob:wmmse_min} is a convex problem and, as such,
efficiently solvable.  The complete centralized algorithm is outlined in
Algorithm~\ref{alg:wmmse}. As shown in~\cite{Shi-Razaviyayn-Luo-He-11}, the
successive approximation algorithm provides monotonic convergence of the
objective function and convergence to a local stationary point of the original
problem~\eqref{prob:wsrmax}. Alternatively, we could have also applied extended
\ac{WSRMax} techniques such as the ones proposed
in~\cite{Kaleva-Tolli-Juntti-TSP16}, where the authors propose an \ac{SCA}
approach with improved rate of convergence.

\begin{algorithm}
    \caption{Centralized \ac{WMMSE} algorithm.}
\label{alg:wmmse}
    
    \begin{algorithmic}[1]
        \STATE{Initialize feasible $\M{m}_{b,k,l}\ \forall\ (b,k,l)$ and $n =
            1$.}
        \REPEAT{}
        \STATE{Generate the \ac{MMSE} receivers $\M{u}_{k,l}\ \forall\ (k,l)$
            from~\eqref{eq:lmmse}.}
        \STATE{Compute the \ac{MSE} $\epsilon_{k,l}^{(n)}\ \forall\ (k,l)$
            from~\eqref{eq:mse}.}
        \STATE{Set the weights $w_{k,l}^{(n)}\ \forall\ (k,l)$
            from~\eqref{eq:mse_weights}.}
        \STATE{Solve the precoders $\M{m}_{b,k,l}\ \forall\ (b,k,l)$
            from~\eqref{prob:wmmse_min}.}
        \STATE{Set $n = n + 1$.}
        \UNTIL{Desired level of convergence has been reached.}
    \end{algorithmic}
\end{algorithm}


\section{Stream Specific Estimation based Beamforming}
\label{sec:sse}

In this section, we propose decentralized \ac{JP} transceiver design
for~\eqref{prob:mse_min}. We assume perfect pilot estimation, i.e., we do not
have to consider the pilot estimation error or noise. Essentially, all \acp{UE}
are assigned orthogonal system-wide pilot training sequences. This allows the
\acp{BS} and \acp{UE}  to estimate the stream specific pilots without having to
consider interference from overlapping pilot sequences. The issues of pilot
contamination and non-orthogonal pilots are considered in
Section~\ref{sec:direct_estimation}. The beamformer signaling relies heavily on
the channel reciprocity of \ac{TDD}. For more information on precoded pilot
signaling see~\cite{Komulainen-Tolli-Juntti-13}.

In~\cite{Shi-Razaviyayn-Luo-He-11} and~\cite{Komulainen-Tolli-Juntti-13}, it was
shown that \ac{CB} using the \ac{WMMSE} algorithm has inherently decoupled
interference processing. As such, it can be easily decentralized with low
signaling overhead. However, the \ac{JP} transmit beamformer design
in~\eqref{prob:wmmse_min} is coupled between the \acp{BS} due to the coherent
signal reception, which prevents us from directly applying same decentralized
processing method. In the sequel, we propose different approaches for
decentralized \ac{JP}.

\subsection{Best Response}
\label{sec:decentralized_best_response}

The \ac{BR} design employs the parallel optimization scheme proposed
in~\cite{Scutari-Facchinei-Song-Palomar-Pang-14} to decentralize the beamformer
design. This parallel framework is based on solving the beamformers locally in
each \ac{BS}, while assuming that the coupling cooperating \acp{BS} keep their
transmitters fixed. Since each BS relies only on the knowledge of the coupled
transmissions from the previous iteration, the beamforming problem becomes
decoupled.  It was shown in~\cite{Scutari-Facchinei-Song-Palomar-Pang-14} that,
if the local problems are strongly convex, the beamformer updates can be made
monotonic with respect to the original \ac{WSRMax} objective function. Note that
the strong convexity of~\eqref{prob:wmmse_min} follows straightforwardly from
the strong convexity of the individual \ac{MSE} functions~\eqref{eq:mse}.

We start by considering the transmit beamformer design for \ac{BS} $b$, while
assuming that the transmission from the other \acp{BS} is fixed. Keeping this in
mind, the transmit beamformer design in~\eqref{prob:wmmse_min}, in iteration
$n$, can be reformulated as
\begin{equation}
\label{prob:wmmse_min_decoupled}
    \begin{array}{rl} 
        \underset{\substack{\M{m}_{b,k,l}}}{\min.} &  
            \ds 
            \sum_{k \in \mathcal{C}_b}\sum_{l = 1}^{L_k} 
                w^{(n)}_{k,l}\bar{\epsilon}_{b,k,l} + 
            \sum_{k \in \mathcal{K} \setminus \mathcal{C}_b}\sum_{l = 1}^{L_k} 
                w^{(n)}_{k,l}\hat{\epsilon}_{b,k,l}
    \\
    \mathrm{s.\ t.} 
        & \ds \sum_{k \in \mathcal{C}_b}\sum_{l = 1}^{L_k} 
                \|\M{m}_{b,k,l}\|^2 \leq P_b
    \text{,}
\end{array}
\end{equation}
where the \ac{MSE} for the $l^\text{th}$ stream of user $k$ is given as 
\begin{equation}
\begin{array}{rl}
\label{eq:mse_fixed}
    \ds \bar{\epsilon}_{b,k,l} =  & \ds
        |\M{u}_{k,l}\herm\M{H}_{b,k} \M{m}_{b,k,l} + c_{b,k,l}^{k,l} - 1|^2 + 
        \|\M{u}_{k,l}\|^2\sigma_k^2 +
    \\ & \ds
        \hspace{-0.1cm}
        \sum_{i \in \mathcal{C}_b}
        \sum_{\substack{j = 1,\\ (i,j) \neq (k,l)}}^{L_i}
            |\M{u}_{k,l}\herm\M{H}_{b,k} \M{m}_{b,i,j} + c_{b,k,l}^{i,j}|^2
    \text{,}
\end{array}
\end{equation}
and the interfering \ac{MSE} is
\begin{math} 
    \hat{\epsilon}_{b,k,l} =
    \sum_{i \in \mathcal{C}_b}\sum_{j = 1}^{L_i}
    |\M{u}_{k,l}\herm\M{H}_{b,k} \M{m}_{b,i,j} + c_{b,k,l}^{i,j}|^2
\text{.}
\end{math}
Here, the fixed terms (cooperating transmit beamformers) are
\begin{equation}
\label{eq:fixed_terms}
    c_{b,k,l}^{i,j} = \sum_{r \in \mathcal{B}_k \setminus \{b\}} 
                        \M{u}_{k,l}\herm\M{H}_{r,k}\M{m}^{(n)}_{r,i,j}
\text{,}
\end{equation}
where $i,j$ denote the transmit beamformer for the $j^{th}$ stream of user $i$
and $k,l$ denote the receiving user $k$ over stream $l$.

The monotonic convergence can be guaranteed by imposing a regulation step
after~\eqref{eq:mse_fixed}. For a small enough step-size $\alpha > 0$, the
update regulation is performed as
\begin{equation}
\label{eq:beamformer_update}
    \M{m}^{(n+1)}_{b,k,l} = \M{m}^{(n)}_{b,k,l} + \alpha\left(
        \M{m}^*_{b,k,l} - \M{m}^{(n)}_{b,k,l}
    \right)\ \forall\ (b,k,l)
\text{,}
\end{equation}
where $\M{m}^*_{b,k,l}\ \forall\ (b,k,l)$ is the optimal solution
to~\eqref{prob:wmmse_min}. For further details on the convergence properties and
step-size selection see~\cite{Scutari-Facchinei-Song-Palomar-Pang-14}. For
constant channels, convergent $\alpha$ can be analytically bounded with respect
to the Lipschitz constant of the
objective~\cite{Scutari-Facchinei-Song-Palomar-Pang-14}.

Similar to~\cite{Shi-Razaviyayn-Luo-He-11}, we can derive a closed form solution
for the transmit beamformers by evaluating the \ac{KKT} conditions
of~\eqref{prob:wmmse_min_decoupled}. This gives us the beamformers in form
\begin{equation}
\label{eq:transmit_beamformer}
    \M{m}_{b,k,l} = \M{C}_b^{-1} \M{p}_{k,l},
\end{equation}
where the transmit covariance matrix is given as
\begin{equation}
\label{eq:transmit_covariance}
    \M{C}_b = \sum_{i = 1}^K\sum_{j = 1}^{L_i}
        \M{H}_{b,i}\herm\M{u}_{i,j}w_{i,j}^{(n)}\M{u}_{i,j}\herm\M{H}_{b,i}  + 
        \M{I}\nu_b
\end{equation}
and
\begin{equation}
    \M{p}_{k,l} = \M{H}_{b,k}\herm\M{u}_{k,l} w_{k,l}^{(n)} + 
        \sum_{i \in \mathcal{C}_b}\sum_{j = 1}^{L_i} 
            \M{H}_{b,i}\herm\M{u}_{i,j} w_{i,j}^{(n)} c_{b,i,j}^{k,l}
\text{.}
\end{equation}
The optimal transmit beamformers can be determined
from~\eqref{eq:transmit_beamformer} by bisection search over $\nu_b$ in such a
way that the transmit power constraints $\sum_{k \in \mathcal{C}_b}\sum_{l =
1}^{L_k} \|\M{m}_{b,k,l}\|^2 \leq P_b$ hold.  Note that if $\sum_{k \in
\mathcal{C}_b}\sum_{l = 1}^{L_k} \|\M{m}_{b,k,l}\|^2 < P_b$ for $\nu_b = 0$,
then this is the optimal solution. Furthermore, the dimensions
of~\eqref{eq:transmit_covariance} depend only on the number of antennas in
\ac{BS} $b$ ($N_\text{T}$) and not the dimensions of the joint beamformer
($|\mathcal{B}_k|N_\text{T}, k = 1,\ldots,K$). This is {\it a considerable
reduction in terms of computational complexity}, when compared to solving the
joint beamformers directly from~\eqref{prob:wmmse_min} (involves inversion of
matrices with dimension $|\mathcal{B}_k|N_\text{T}, k = 1,\ldots,K$).
Considering that bisection converges
fast~\cite{Boyd-Vandenberghe-04},~\eqref{eq:transmit_beamformer} gives us a low
complexity way to solve the beamformers without having to resort on general
convex solvers.  Finally, the decentralized beamformer design has been
summarized in Algorithm~\ref{alg:wmmse_decentralized}.
\begin{algorithm}
    \caption{Decentralized \ac{BR} \ac{WMMSE} algorithm.}
\label{alg:wmmse_decentralized}

    \begin{algorithmic}[1]
        \STATE{Initialize feasible $\M{m}_{b,k,l}\ \forall\ (b,k,l)$ and $n =
            1$.}
        \REPEAT{}
        \STATE{{\it \ac{UE}\/}: Generate the \ac{MMSE} receivers $\M{u}_{k,l}\
            \forall\ (k,l)$ from~\eqref{eq:lmmse}.}
        \STATE{{\it \ac{UE}\/}: Compute the \ac{MSE} $\epsilon_{k,l}^{(n)}\
            \forall\ (k,l)$ from~\eqref{eq:mse}.}
        \STATE{{\it \ac{UE}\/}: Set the weights $w_{k,l}^{(n)}\ \forall\ (k,l)$
            from~\eqref{eq:mse_weights}.}
        \STATE{{\it \ac{BS}\/}: Solve the precoders $\M{m}_{b,k,l}\ \forall\
            (b,k,l)$ from~\eqref{eq:transmit_beamformer}.}
        \STATE{{\it \ac{BS}\/}: Update the next iteration precoders according
            to~\eqref{eq:beamformer_update}.}
        \STATE{Set $n = n + 1$.}
        \UNTIL{Desired level of convergence has been reached.}
    \end{algorithmic}
\end{algorithm}

\subsubsection*{Signaling Requirements}

Solving the \ac{MMSE} receive beamformers requires only the knowledge of the
precoded downlink channels. That is, each \ac{UE} needs to know
$\M{H}_{b,k}\M{m}_{b,i,z}\ \forall\ (b,i,z)$. On the other hand, solving the
transmit beamformers requires the knowledge of the fixed terms $c_{b,k,l}^{i,j}$
and \ac{MSE} weights $w_{k,l}\ \forall\ (k,l)$ from~\eqref{eq:mse_weights} need
to be exchanged for each frame among the serving \acp{BS}. This can be done
either by using a separate feedback channel from the terminals or over the
backhaul (solely between the \acp{BS}).

Using only the backhaul, each \ac{BS} $b\in\mathcal{B}_i$ can estimate the
corresponding $c_{b,k,l}^{i,j}$ based on the effective \ac{DL} channel
$\mathbf{u}_{k,l}\herm\mathbf{H}_{b,k}$ and the previous iteration precoder
$\mathbf{m}^{(n)}_{b,i,j}$. Then, the terms $c_{b,k,l}^{i,j}$ are distributed
over the backhaul to the cooperating \acp{BS} that form
complete~\eqref{eq:fixed_terms} by summing the corresponding terms. The backhaul
signaling scheme is efficient in the sense that it does not require additional
signaling or estimation effort from the user terminals.

Alternatively, the terminals can estimate the combined signals
\begin{equation}
\label{eq:combined_fixed_terms}
    c_{k,l}^{i,j} = \sum_{r \in \mathcal{B}_k} 
        \mathbf{u}_{k,l}\herm\mathbf{H}_{r,k,l}\mathbf{m}^{(n)}_{r,i,j}
\end{equation}
from the precoded \ac{DL} pilots ($\mathbf{H}_{b,k}\mathbf{m}_{b,i,j}\ \forall\
(b,i,j)$). The combined signals~\eqref{eq:combined_fixed_terms} are then
distributed to the \acp{BS} over a feedback channel. Each BS $b$ can then form
$c_{b,k,l}^{i,j}$ by subtracting its own part
from~\eqref{eq:combined_fixed_terms}, i.e., $c_{b,k,l}^{i,j} = c_{k,l}^{i,j} -
\mathbf{u}_{k,l}\herm\mathbf{H}_{b,k,l}\mathbf{m}^{(n)}_{b,i,j}$.  This will
somewhat increase the computational burden of the terminals, as the users need
to estimate also the precoded pilot signals from the interfering sources. Note
that this does not require additional \ac{DL} pilot resources as the pilots are
assumed to be orthogonal for each stream in any case. The signaling schemes can
be summarized as
\begin{enumerate}
    \item Backhaul offloading for the fixed terms~\eqref{eq:fixed_terms} can be
        used to reduce the signaling requirements of the user terminals. In this
        case, each BS reports their corresponding part of $c_{b,k,l}^{i,j}$ over
        the backhaul to the cooperating \acp{BS}. The effective DL channels are
        still obtained from the UL pilots.
    \item Feedback channel signaling, where users estimate the sum received
        signals\eqref{eq:combined_fixed_terms} and broadcast them over a
        feedback channel to the serving \acp{BS}. Each user reports separately
        the intended signal and all of the interfering streams.
    \item If global CSI is exchanged, every BS can solve the complete global
        problem locally.
\end{enumerate}
It should be noted that, with efficient clustering of the serving \acp{BS}, the
signaling overhead can be significantly decreased. \acp{BS} far from the users
do not contribute meaningful gain to the joint processing, and can, as such, be
neglected from the serving sets. An example of the signaling requirements of the
fixed terms in~\eqref{eq:fixed_terms} for 7-cell system with $K = 49$ users in
total is given in Table~\ref{tbl:signaling}. This is a worst case scenario,
where each user is assumed to be served with the maximum number $L_k = 2$
streams and all \acp{BS} serve every user coherently. 





\subsection{Alternating Direction Method Multipliers}
\label{sec:decentralized_admm}
While providing a low complexity implementation for
solving~\eqref{prob:wmmse_min}, the best response approach in
Section~\ref{sec:decentralized_best_response} is sensitive to proper step-size
selection.  A more robust alternative can be achieved by using the \ac{ADMM}
approach, which has been shown to provide efficient decomposition and good
convergence properties for various types of
problems~\cite{Boyd-Parikh-Chu-Peleato-Eckstein-10}. \Ac{ADMM} can be seen as an
extension for dual decomposition based techniques with improved convergence
properties.

The starting point for the dual based decomposition design, is to gather the
coupling variables into locally and globally updated
components~\cite{Boyd-EE364b-PrimDualDecomp-07}. To this end,  we introduce
auxiliary variables $s_{k,l,b,i,z}$ and constraints to denote the received
symbol over the $l^\text{th}$ spatial stream of user $k$ from \ac{BS} $b$, which
is intended for stream $z$ of user $i$. These variables are imposed in form of
consensus constraints
\begin{equation}
\label{eq:consensus}
    s_{k,l,b,i,z} = \sqrt{w_{k,l}}\M{u}_{k,l}\herm\M{H}_{b,k}\M{m}_{b,i,z}
\ \forall\ (k,l,i,z), b \in \mathcal{B}_i
\text{.}
\end{equation}
The dual variables (Lagrangian multipliers) related to~\eqref{eq:consensus} are
then denoted as $\lambda_{k,l,b,i,z}$.  The principal idea in \ac{ADMM} is to
alternate the updates of variables $s_{k,l,b,i,z}$ and $\M{m}_{b,i,z}$ along
with the dual variables $\lambda_{k,l,b,i,z}$ of~\eqref{eq:consensus} while
keeping the others fixed.

Now, to separate the updates, we use the partial Lagrangian relaxation of the
constraints~\eqref{eq:consensus}.  Additionally, we impose penalty norms for the
constraint violation, which are used to enforce the consensus
in~\eqref{eq:consensus} and improve the rate of convergence. These penalty terms
are given as
\begin{equation}
\label{eq:admm_penalty}
    \Theta_{k,l} = 
        \sum_{i = 1}^K\sum_{z = 1}^{L_i}\sum_{b\in\mathcal{B}_i}
            \frac{\rho}{2}
                |
                    \sqrt{w_{k,l}}\M{u}_{k,l}\herm\M{H}_{b,k}\M{m}_{b,i,z} - 
                    s_{k,l,b,i,z} + \lambda_{k,l,b,i,z}
                |^2
    \text{,}
\end{equation}
where parameter $\rho$ is adjusted to determine the degree of enforcement for
constraints~\eqref{eq:consensus}.  Note that the dual
variables~$\lambda_{k,l,b,i,z}$ in~\eqref{eq:summse_admm_lag_0} are scaled so
that they can be incorporated into the penalty norms. For a detailed discussion
on $\rho$ balancing and scaled dual variables,
see~\cite{Boyd-Parikh-Chu-Peleato-Eckstein-10}. 

Similarly to the \ac{BR} approach in
Section~\eqref{sec:decentralized_best_response}, we can combine the transmission
over the \ac{JP} clusters  as 
\begin{equation}
\label{eq:admm_sum_interference}
    \bar{s}_{k,l,i,z} = \sum_{b \in\mathcal{B}_i} 
                            s_{k,l,b,i,z} \ \forall\ (k,l,i,z) 
    \text{.}
\end{equation}
This denotes the coherent transmission of transmission for the $z^\text{th}$
stream of user $i$ perceived over the $l^\text{th}$ stream of user $k$.  It
should also be noted that $s_{k,l,b,i,z}\ \forall\ (k,l,b,i,z)$ and
$\bar{s}_{k,l,i,z}\ \forall\ (k,l,i,z)$ are complex variables
in~\eqref{eq:summse_admm_lag_0}. For the notional convenience, also,
$\lambda_{k,l,b,i,z}$ are complex number.

Now, using~\eqref{eq:consensus} and~\eqref{eq:admm_sum_interference} we can
rewrite the \ac{MSE} expression for stream $l$ of user $k$ as
\begin{equation}
\label{eq:admm_mse}
    \tilde{\epsilon}_{k,l} = 
        w_{k,l} - 
        2\sum_{b \in \mathcal{B}_k}
            \Re\{w_{k,l}\M{u}_{k,l}\herm\M{H}_{b,k}\M{m}_{b,k,l}\} + 
        \sum_{i = 1}^K \sum_{z = 1}^{L_i} |\bar{s}_{k,l,i,z}|^2
    \text{.}
\end{equation} 
With the help of~\eqref{eq:admm_mse}, the primal optimization problem, for fixed
$\lambda_{k, l, b, i, z}$ becomes
\begin{equation}
\label{eq:summse_admm_lag_0}
    \begin{array}{ll}
        \underset{\substack{s_{k,l,b,i,z}, \\ \bar{s}_{k,l,i,z},\\ 
                            \M{m}_{b,k,l}}}{\min.} 
        & \ds \sum_{k = 1}^K \sum_{l = 1}^{L_k}
        \left(\tilde{\epsilon}_{k,l} + \Theta_{k,l}\right)
    \\
        \mathrm{s.\ t.} 
        & \ds~\eqref{eq:admm_sum_interference}, \\
        & \ds \sum_{k \in \mathcal{C}_b}
            \|\M{m}_{b,k}\|^2 \leq P_b,\ b = 1,\ldots,B
        \text{.}
    \end{array}
\end{equation}
The dual update is then given as
\begin{equation}
\label{eq:dual_update_simple}
    \lambda_{k,l,b,i,z}^{(n+1)} = \lambda_{k,l,b,i,z}^{(n)} + 
        \sqrt{w_{k,l}}\M{u}_{k,l}\herm\M{H}_{j,k}\M{m}_{b,i,z}^{(n+1)} - 
        s_{k,l,b,i,z}^{(n+1)}
    \text{.}
\end{equation}

Decentralized solution for~\eqref{eq:summse_admm_lag_0} would still require
exchanging all $s_{k,l,b,i,z}\ \forall\ (k,l,b,i,z)$ within the serving set
$\mathcal{B}_i$. Also, each UE $k$ would need to be able to separate individual
effective channels $\M{H}_{b,k}\M{m}_{b,i,z}\ \forall\ i \in \mathcal{K}, z =
1,\ldots, L_i$, which is intractable as it would require orthogonal pilot
signaling within each cooperating set of \acp{BS} $\mathcal{B}_k, k \in
\mathcal{K}$.

Problem~\eqref{eq:summse_admm_lag_0} can be further simplified, in such a way
that the problem is coupled only via the summed
signals~\eqref{eq:admm_sum_interference} instead of the individual
$s_{k,l,b,i,z}\ \forall\ (k,l,b,i,z)$. The reformulation is quite technical and
has, thus, been provided in Appendix~\ref{ap:admm_reformulation}. In the end, we
can solve $\bar{s}_{k,l,i,z}\ \forall\ (k,l,i,z)$
from~\eqref{eq:summse_admm_lagr}, for fixed beamformers $\M{m}_{b,k,l}^{(n+1)}\
\forall\ (b,k,l)$ and dual variables $\bar{\lambda}_{k,l,i,z}^{(n)}$, as
\begin{align}
\label{eq:admm_lin_eq_s1}
    \bar{s}_{k,l,i,z}^{(n+1)} &= 
        \frac{\rho}{1 + \rho}
        \left(
            \sum_{b\in\mathcal{B}_i}
                \sqrt{w}_{k,l}\M{u}_{k,l}\herm\M{H}_{b,k}\M{m}_{b,i,z}^{(n+1)} + 
                \bar{\lambda}_{k,l,i,z}^{(n)}
        \right)
        \text{.}
\end{align}
The dual variable update is then given as
\begin{equation}
\label{eq:dual_update}
    \bar{\lambda}_{k,l,i,z}^{(n+1)} = \bar{\lambda}_{k,l,i,z}^{(n)} + 
    \sum_{g\in\mathcal{B}_i} 
        \sqrt{w_{k,l}}\M{u}_{k,l}\herm\M{H}_{j,k}\M{m}_{g,i,z}^{(n+1)} - 
        \bar{s}_{k,l,i,z}^{(n+1)}
    \text{.}
\end{equation}

Now, with~\eqref{eq:cohupdate_single} and dual update~\eqref{eq:dual_update},
Problem~\eqref{eq:summse_admm_lag_0} can be formulated as 
\begin{equation}
\label{eq:summse_admm_lagr}
    \begin{array}{ll}
        \underset{\substack{\bar{s}_{k,l,i,z},\\ \M{m}_{b,k,l}}}{\min.} 
        & \ds \sum_{k = 1}^K \sum_{l = 1}^{L_k}
            \left(
                \tilde{\epsilon}_{k,l} + 
                \sum_{i = 1}^K\sum_{z = 1}^{L_i}\sum_{j\in\mathcal{B}_i}
                    \Psi_{k,l,i,z}
            \right)
        \\
        \mathrm{s.\ t.} 
        & \ds \sum_{k \in \mathcal{C}_b}\sum_{l = 1}^{L_k}
                \|\M{m}_{b,k,l}\|^2 \leq P_b,\ b = 1,\ldots,B
        \text{,}
    \end{array}
\end{equation}
where 
\begin{equation}
    \Psi_{k,l,b,i,z} =  
        \rho |\sqrt{w_{k,l}}\M{u}_{k,l}\herm\M{H}_{j,k}\M{m}_{b,i,z} - 
              q_{b,k,l,i,z}|^2
\end{equation}
and
\begin{equation}
    q_{b,k,l,i,z} = 
        \sum_{g\in\mathcal{B}_i\setminus \{b\}} 
            \sqrt{w_{k,l}}\M{u}_{k,l}\herm\M{H}_{j,k}\M{m}_{g,i,z} - 
            \bar{s}_{k,l,i,z}^{(n)} + \bar{\lambda}_{k,l,i,z}
    \text{.}
\end{equation}
Similarly to the \ac{BR} design (Section~\ref{sec:decentralized_best_response}),
the transmit beamformers can be solved from~\eqref{eq:summse_admm_lagr} by
closed form bisection search. From the first order optimality condition, for
fixed $\bar{s}_{k,l,i,z}$, we have the beamformers in form
\begin{equation}
\label{eq:admm_closed_form_beamformers}
    \M{m}_{b,k,l} = \M{C}_{b}^{-1}\left(\M{H}_{b,k}\herm\M{u}_{k,l}w_{k,l} + 
                    \M{f}_{b,k,l}\right)
    \text{,}
\end{equation}
where 
\begin{equation}
\label{eq:admm_closed_form_covariance}
    \M{C}_{b} = 
        \sum_{i = 1}^K\sum_{j = 1}^{L_i}
            \M{H}_{b,i}\herm\M{u}_{i,j}\rho w_{i,j}\M{u}_{i,j}\herm\M{H}_{b,i} +
            \M{I}\nu_b
    \text{.}
\end{equation}
and
\begin{equation}
\label{eq:admm_intf}
    \M{f}_{b,k,l} = 
        \sum_{i = 1}^K\sum_{z = 1}^{L_i}
            \mathbf{H}_{b,i}\herm\M{u}_{i,z}\sqrt{w_{i,z}}\rho q_{b,k,l,i,z}
            \text{.}
\end{equation}
The transmit beamformers are then determined
from~\eqref{eq:admm_closed_form_beamformers} by bisection search over $\nu_b$ so
that that the transmit power constraints $\sum_{k \in \mathcal{C}_b}\sum_{l =
1}^{L_k} \|\M{m}_{b,k,l}\|^2 \leq P_b$ hold. 

It can be observed that, both, the update~\eqref{eq:admm_lin_eq_s1} and dual
update~\eqref{eq:dual_update} can be managed locally at each BS $b$, if the
averaged coherently received signals 
\begin{math}
    \sum_{g\in\mathcal{B}_i} 
        \sqrt{w_{k,l}}\M{u}_{k,l}\herm\M{H}_{j,k}\M{m}_{g,i,z}^n
\end{math}
are known within the joint processing clusters.  The outline of the \ac{ADMM}
algorithm is given in Algorithm~\ref{alg:admm}.

\subsubsection*{Signaling Requirements}
We can see from~\eqref{eq:admm_closed_form_beamformers} that the beamformer
structure is similarly to the \ac{BR} design~\eqref{eq:transmit_beamformer}.
This also leads to similar signaling requirements. In fact, the signaling
requirements turnout to be equivalent. The proposed \ac{ADMM} method also
requires the exchange of the received signal from the cooperating \acp{BS} for,
both, the intended signal and interference in order to be able to perform the
update in~\eqref{eq:admm_lin_eq_s1}.

\subsubsection*{Convergence}
Due to lack of space, detailed convergence analysis is neglected. Proof of
convergence for the \ac{ADMM} method for convex problems can be found
from~\cite{Boyd-Parikh-Chu-Peleato-Eckstein-10}. These results can be applied to
the beamformer convergence for fixed receive filters.  However, as the
\ac{WMMSE} transceiver design is, in general, non-convex
(see~\cite{Luo-Davidson-Giannakis-Wong-04} for convexity conditions), the
receive filter update~\eqref{eq:lmmse} requires extended analysis. Rough
convergence conditions can be derived by noting that the receive filter update
strictly improves the objective value. Now, such conditions for $\rho$ can
derived that, after each full iteration, Algorithm~\ref{alg:admm} moves towards
a stationary point of~\eqref{prob:wsrmax}. As for the recent developments on
solving non-convex \ac{ADMM} see~\cite{Hong-Luo-Razaviyayn-14}.

\begin{algorithm}
	\caption{\ac{ADMM} algorithm for \ac{WSRMax}}
\label{alg:admm}
	
	\begin{algorithmic}[1]
        \STATE{{\it \ac{UE}\/}: Initialize the \ac{MMSE} receive filters
            $\M{u}_{k,l}\ \forall\ (k, l)$.}
        \STATE{{\it \ac{BS}\/}: Initialize the variables $\bar{s}_{k,l,i,z}^n =
            0$ and dual variables $\bar{\lambda}_{k,l,i,z}^n = 0$ for all
            $(k,l,i,z)$.}
		\REPEAT{}
        \STATE{{\it \ac{BS}\/}: Update the local beamformers
            from~\eqref{eq:summse_admm_lagr}.}          
        \STATE{{\it \ac{BS}\/}: Locally update the variables $\bar{s}_{k,l,i,z}$
            and dual variables $\bar{\lambda}_{k,l,i,z}$
            from~\eqref{eq:admm_lin_eq_s1} and~\eqref{eq:dual_update}.}
        \STATE{{\it \ac{UE}\/}: Update the receive filters $\M{u}_{k,l}\
            \forall\ (k,l)$ from~\eqref{eq:lmmse}.}
        \STATE{{\it UE\/}: Compute the \ac{MSE} $\epsilon_{k,l}^{(n)}\ \forall\
            (k,l)$ from~\eqref{eq:mse}.}
        \STATE{{\it UE\/}: Set the weights $w_{k,l}^{(n)}\ \forall\ (k,l)$
            from~\eqref{eq:mse_weights}.}
		\UNTIL{Desired level of convergence has been reached.}
	\end{algorithmic}
\end{algorithm}

\subsection{Stochastic Gradient Descent}

The best response and \ac{ADMM} based decentralized \ac{JP} techniques have
attractive convergence properties. As a low complexity alternative to the
aforementioned approaches, we propose a \ac{SG} method. This method is based on
updating the transmit beamformers, in each iteration, solely into the direction
of the objective gradient, which greatly simplifies the transceiver processing.

For notational convenience, we begin by denoting the  weighted effective
channels from \ac{BS} $b$ over the $l^\text{th}$ stream of \ac{UE} $k$ as 
\begin{equation}
\label{eq:effective_dl_channel}
    \tilde{\M{y}}_{b,k,l}\herm = \sqrt{w}_{k,l}\M{u}_{k,l}\herm\M{H}_{b,k} 
    \text{.}
\end{equation}
Similar to~\eqref{eq:admm_mse}, the weighted \ac{MSE} terms can be written with
the help of~\eqref{eq:effective_dl_channel} as
\begin{equation}
\begin{array}{r}
	\ds \tilde{\epsilon}_{k,l} = w_{k,l} 
		- 2\sum_{b \in \mathcal{B}_k}
            \Re\{\sqrt{w}_{k,l}\tilde{\M{y}}_{b,k,l}\herm\M{m}_{b,k,l}\}
		+ \\ 
		\ds \sum_{i = 1}^K \sum_{z = 1}^{L_i} 
			|\sum_{b \in \mathcal{B}_i} 
                \tilde{\M{y}}_{b,k,l}\herm \M{m}_{b,i,z}|^2
	\text{.}
\end{array}
\end{equation} 
Next, we reformulate the \ac{WMMSE} objective of~\eqref{prob:wmmse_min}
equivalently as\footnote{The constants terms have been neglected as they do not
contribute to the optimal solution.}
\begin{equation}
\label{eq:sg_objective}
\begin{array}{r}
	\ds \sum_{k=1}^K\sum_{l=1}^{L_k}\left(
		\sum_{j = 1}^K \sum_{z = 1}^{L_i}
			|\sum_{b \in \mathcal{B}_j} 
                \tilde{\M{y}}_{b,k,l}\herm \M{m}_{b,j,z}|^2
		 -
		\right. \\ 
	\ds \left.  
		\hspace{1cm}
		2\sum_{b \in \mathcal{B}_k}
            \Re\{\sqrt{w}_{k,l}\M{y}_{b,k,l}\herm\M{m}_{b,k,l}\}
		\right)
	\text{.}
\end{array}
\end{equation}
The gradient of~\eqref{eq:sg_objective} in terms of $\M{m}_{b,k,l}$ can be given
as 
\begin{equation}
\label{eq:sg_grad}
\M{G}_{b,k,l} = 
    2\sum_{i=1}^K \sum_{z=1}^{L_j}
        \tilde{\M{y}}_{b,i,z}
        \sum_{g\in\mathcal{B}_k}\tilde{\M{y}}_{g,i,z}\herm\M{m}_{g,k,l}
    - 2\M{H}_{b,k} \M{u}_{k,l}\herm w_{k,l}
\text{.}
\end{equation}
Note that the gradient expression~\eqref{eq:sg_grad} does not become fully
decoupled among the \acp{BS} due to the terms
\begin{math}
   \sum_{g\in\mathcal{B}_i}\tilde{\M{y}}_{g,i,z}\herm\M{m}_{g,k,l}
    \text{.}
\end{math}
The \ac{SG} update in the direction of the gradient, in iteration $n$, is given as 
\begin{equation}
\label{eq:sg_update}
    \M{m}_{b,k,l}^{(n+1)} = \M{m}_{b,k,l}^{(n)} - 
                            \alpha \M{G}^{(n)}_{b,k,l} \ \forall\ (b, k, l)
    \text{.}
\end{equation}
It is evident that~\eqref{eq:sg_update} can be independently performed at each
\acp{BS} $b$ if
\begin{equation}
\label{eq:sg_reg_info}
    \tilde{\M{y}}_{\bar{b},k}\herm\M{m}_{\bar{b},k,l}^{(n)} 
        \forall\ \bar{b} \notin \mathcal{B}_k
\end{equation}
are made available. However,~\eqref{eq:sg_update} alone is not sufficient for
accurate beam coordination with \ac{JP} as it does not take into account the
power control. That is,~\eqref{eq:sg_update} may lead to a solution, where the
available power budget ($P_b$) is exceed. To address this problem, we propose
two approaches for the power control.

\subsubsection{Feasible projection}
A straightforward approach for power control is to simply scale the beamformers
to meet the power constraints. That is, if $\sum_{k \in \mathcal{C}_b}^K\sum_{l
= 1}^{L_k} \|\M{m}_{b,k,l}\|^2 > P_b$ for some $b = 1,\ldots,B$, the
corresponding \ac{BS} scales the beamformers by $\frac{\sqrt{P_b}}{\sum_{k \in
\mathcal{C}_b}^K\sum_{l = 1}^{L_k} \|\M{m}_{b,k,l}\|}$. The problem here is that
the scaling is not global in the sense that each \ac{BS} use different scaling.
This also changes the direction of the beamformer, which may have detrimental
impact on the performance.

\subsubsection{Dual decomposition}
\label{sec:cse_sg_dual}
More sophisticated and better performing power control can be achieved by
employing the dual decomposition technique to steer the beamformer updates
in~\eqref{eq:sg_update} towards the feasible set.  Using the dual decomposition,
we get the augmented Lagrangian for~\eqref{prob:wmmse_min} in form
\begin{equation}
\label{eq:sg_lagrangian}
    \ds~\eqref{eq:sg_objective} + 
    \sum_{b = 1}^B
        \nu_b\left(
            \sum_{k \in \mathcal{C}_b}^K\sum_{l = 1}^{L_k} 
                \|\M{m}_{b,k,l}\|^2 - P_b
        \right)
    \text{,}
\end{equation}
where $\nu_b, b = 1, \ldots, B$ are the dual variable corresponding to the power
constraints. Taking the gradient of~\eqref{eq:sg_lagrangian}, we get 
\begin{equation}
\label{eq:sg_dual_grad}
    \bar{\M{G}}^{(n)}_{b,k,l} = \M{G}^{(n)}_{b,k,l} + \nu_b \M{m}_{b,k,l}
    \text{,}
\end{equation}
It follows from the gradient update approach, that the \ac{SG} step becomes
\begin{equation}
\label{eq:sg_dual_update}
    \M{m}_{b,k,l}^{(n+1)} = \M{m}_{b,k,l}^{(n)} - 
                            \alpha \bar{\M{G}}^{(n)}_{b,k,l}
    \text{.}
\end{equation}
Now, to steer the beamformer updates towards feasible power levels, after
each~\eqref{eq:sg_dual_update}, we update the dual variables as 
\begin{equation}
\label{eq:sg_dual_variable}
    \nu_b^{(n+1)} = 
    \max\left(0, 
        \nu_b^{(n)} + 
        \beta\left(
            P_b - 
            \sum_{k \in \mathcal{C}_b}^K\sum_{l = 1}^{L_k}
                \|\M{m}_{b,k,l}^{(n)}\|^2
        \right)
    \right)
    \text{,}
\end{equation}
where $\beta$ is a sufficiently small step size.  The decentralized \ac{SG}
algorithm is outlined in Algorithm~\ref{alg:sg}.

\subsubsection*{Regularized updates}
As the \ac{SG} update are based solely on the currently available gradient,
these updates can be, in some cases, overly aggressive. Step-size normalization
is the most straightforward way to regularize the absolute step size. Normalized
step size is then given as
\begin{equation}
    \tilde{\alpha}_{k,l} = \frac{\alpha}{\|\M{G}^{(n)}_{k,l}\|^2}
    \text{,}
\end{equation}
where $\M{G}_{k,l}^{(n)}$ is the full gradient vector for $(k,l)$.

Another way to regularize the \ac{SG} updates, is to make the gradient update
more dependent on the previous update direction. In another words this adds
momentum for the general tendency of the update direction. The momentum is
adaptively updated as
\begin{equation}
    \M{M}_{b,k,l}^{(n+1)} = \M{G}_{b,k,l} + \omega\M{M}_{b,k,l}^{(n)} 
    \text{,}
\end{equation}
where $\omega \geq 0$ denotes the momentum magnitude. In principal this is close
to the regularized \ac{BR} update procedure in~\eqref{eq:beamformer_update}.
Finally, the beamformer update becomes
\begin{equation}
\label{eq:sg_momentum_dual_update}
    \M{m}_{b,k,l}^{(n+1)} = \M{m}_{b,k,l}^{(n)} - 
                            \alpha \M{M}^{(n+1)}_{b,k,l}
    \text{.}
\end{equation}
The regularized update routines are particularly helpful in fading channels,
where the gradient of the instantaneous channel realization may not fully
represent the overall fading conditions. This is demonstrated by numerical
examples in Section~\ref{sec:simres}.

\subsubsection*{Signaling Requirements}
When comparing to the methods in Sections~\ref{sec:decentralized_best_response}
and~\ref{sec:decentralized_admm}, the signaling requirements of the \ac{SG}
design are identical. Assuming that \ac{TDD} is employ and the local effective
channels are estimated from the uplink pilots~\cite{Komulainen-Tolli-Juntti-13},
the per-stream \ac{MSE} information and~\eqref{eq:sg_reg_info} need to be
explicitly shared among the \acp{BS}.  The \ac{MSE} sharing has been extensively
studied in~\cite{Komulainen-Tolli-Juntti-13} and can be done roughly in two
ways. Either the \acp{UE} send the \ac{MSE} information as feedback to the
\acp{BS} or the \acp{BS} share their contribution to the individual \ac{MSE}
terms over the backhaul. 

\subsubsection*{Convergence}
While the conventional \ac{SG} method is known to converge with sufficiently
small step size~\cite{Haykin-96}, the proposed method involve the iterative
receive beamformer update and \ac{SCA} of the objective function. Incorporating
these steps to the convergence analysis is out of the scope of this manuscript.
In any case, we can always iterate the \ac{SG} and dual update steps
sufficiently long to guarantee improved objective value, which in turn can be
used to provide simple proof of convergence for the objective. On the other
hand, our simulation results indicate that the algorithm convergences even for
single iteration between each each step (as shown in Algorithm~\ref{alg:sg}).

\begin{algorithm}
	\caption{Stochastic Gradient Ascent.}
\label{alg:sg}
	
	\begin{algorithmic}[1]
        \STATE{Initialize feasible $\M{m}_{b,k,l}\ \forall\ (b,k,l)$ and $n =
            1$.}
		\REPEAT{}
        \STATE{Generate the \ac{MMSE} receivers $\M{u}_{k,l}\ \forall\ (k,l)$
            from~\eqref{eq:lmmse}.}
        \STATE{Compute the \ac{MSE} $\epsilon_{k,l}^{(n)}\ \forall\ (k,l)$
            from~\eqref{eq:mse}.}
        \STATE{Set the weights $w_{k,l}^{(n)}\ \forall\ (k,l)$
            from~\eqref{eq:mse_weights}.}
        \STATE{Update the precoders $\M{m}_{b,k,l}\ \forall\ (b,k,l)$
            from~\eqref{eq:sg_dual_grad}.}
		\STATE{Set $n = n + 1$.}
		\UNTIL{Desired level of convergence has been reached.}
	\end{algorithmic}
\end{algorithm}

\begin{table*}
	\renewcommand{\arraystretch}{1.3}
    \caption{Required amount of information exchange per active data stream in
        7-cell model with $K = 49$, $N_\text{T} = 8$, $N_\text{R} = 2$ and $L_k
        = 2\ \forall\ k = 1,\ldots,K$.}
\label{tbl:signaling}
	\centering
	\begin{tabular}{p{3cm}p{3.5cm}p{3.5cm}p{3.5cm}}
		\toprule
		\bfseries  Signaling Scheme &  
        \bfseries Best Response & 
        \bfseries \Ac{ADMM} & 
        \bfseries \Ac{SG} \\ 
		\midrule
		\midrule
        \bfseries Backhaul offloading (1). Shared symbols between cooperating
            \acp{BS}. & 
		$\sum_{k = 1}^K L_k = 98$ & 
        $\sum_{k = 1}^K L_k = 98$ & 
        $\sum_{k = 1}^K L_k = 98$  
		\\ 
		\midrule
		\bfseries Feedback channel (2). Each \ac{UE} reports to all \acp{BS}. &
		$\sum_{k = 1}^K L_k = 98$ & 
        $\sum_{k = 1}^K L_k = 98$ & 
        $\sum_{k = 1}^K L_k = 98$ 
		\\
		\midrule
		\bfseries Global CSI (3) Shared symbols between cooperating \acp{BS}. & 
		$K N_\text{R}N_\text{T} = 784$ & 
        $K N_\text{R}N_\text{T} = 784$ & 
        $K N_\text{R}N_\text{T} = 784$
		\\
		\bottomrule
	\end{tabular}
\end{table*}

\section{Direct Estimation}
\label{sec:direct_estimation}

In this section, we consider \ac{JP} beamformer design, when the stream specific
pilot estimation may cannot be done accurately. In Section~\ref{sec:sse}, we
basically considered a system, where there are enough pilot resources so that
the stream specific pilots can be allocated orthogonally. We also neglected the
pilot estimation noise, thus, assuming infinite pilot power. This is fairly
common assumption~\cite{Shi-Razaviyayn-Luo-He-11, Komulainen-Tolli-Juntti-13,
Tolli-Pennanen-Komulainen-TWC-11, Bogale-Vandendorpe-11}, and it makes the
beamformer designs somewhat more straightforward.  However, in highly dense
systems, the orthogonal pilot resource allocation may not be tractable, as the
\ac{CSI} estimation also requires knowledge of the pilot sequence used in the
adjacent cells.  Due to the high number of simultaneous transmissions the
orthogonal pilot sequence lengths would increase to be unreasonably large. To
this end, we propose a \ac{DE} of the \ac{MMSE} beamformers, where the pilot
sequence orthogonality can be relaxed. We still allow orthogonal pilots to be
used at least partially to alleviate some of the cross user interference.
Nevertheless, we do not require any particular pilot design over the users. When
considering the pilot design difficulty in dense and heterogeneous networks, one
possibility would be to make the pilots orthogonal only within each \ac{JP}
cluster.

\subsection*{Uplink beamformer estimation}
Let $\M{b}_{k,l} \in \mathbb{C}^{S}$ denote the \ac{UL} pilot training sequence
for the $l^\text{th}$ data stream of \ac{UE} $k = 1,\ldots,K$, where $S$ is the
length of the pilot sequence. Then, the composite of the precoded uplink pilot
training matrices as received at \ac{BS} $b$ is 
\begin{equation}
    \label{eq:uplink_training}
    \M{R}_{b} = \sum_{k = 1}^K \sum_{l = 1}^{L_k}
        \M{H}_{b,k}\herm \M{u}_{k,l} \sqrt{w_{k,l}} \M{b}_{k,l}\herm + \M{N}_b 
    \text{,}
\end{equation}
where $\M{N}_b \in \mathbb{C}^{N_\text{T} \times S}$ is the estimation noise
matrix for all pilot symbols. Similarly, to the \ac{SSE} methods, we employ
precoded training pilots, where the weighted receive beamformers serve as
precoders.

We begin beamformer design by formulating the \ac{LS} estimation objective of
the downlink beamformers. After, which we provide a modification to the \ac{LS}
design, such that the optimization objective matches the \ac{WMMSE}. 

The \ac{LS} objective is given as
\begin{equation}
    \label{eq:direct_ls}
    \begin{array}{l}
        \ds
        \underset{\mathbf{m}_{b,k,l}}{\min.}\quad
            \frac{1}{S} \sum_{(k,l)}
                \|\M{b}_{k,l}\herm - 
                  \sum_{b\in\mathcal{B}_k}\M{m}_{b,k,l}\herm{\M{R}}_b\|^2 = 
            \\
        \ds
            \underset{\mathbf{m}_{b,k,l}}{\min.}\quad
                \sum_{(k,l)}\left(1 -  2\Re\{\sqrt{w_{k,l}}\M{u}_{k,l}\herm 
                    \sum_{b\in\mathcal{B}_k}\M{H}_{b,k}\M{m}_{b,k,l}\} +
                 \right.
            \\
        \ds
            \left. \psi_{k,l} +
            \sum_{b\in\mathcal{B}_k}\M{m}_{b,k,l}\herm 
            \left( 
                \sum_{i = 1}^K\sum_{z = 1}^{L_i} 
                    \M{H}_{b,i}\herm\M{u}_{i,z}w_{i,z}
                    \M{u}_{i,z}\herm\M{H}_{b,i}
            \right)\M{m}_{b,k,l}\right)
            \text{,}
    \end{array}
\end{equation}
where $\psi_{k,l} \geq 0$ defines the estimation error due to the estimation
noise and cross-talk between the pilots.  For fully orthogonal pilots, the error
from pilot cross talk diminishes. If we let the pilot sequence to be orthogonal,
$\M{N}_b \rightarrow \M{0}$ and set $w_{k,l} = 1\ \forall\ (k,l)$,
then~\eqref{eq:direct_ls} becomes equivalent to the sum-\ac{MSE} minimization
objective (see~\eqref{eq:mse}).

Now, the \ac{LS} estimate in~\eqref{eq:direct_ls} can be reformulated as 
\begin{equation}
    \begin{array}{l}
        \ds
        \underset{\mathbf{m}_{b,k,l}}{\min.}\quad
            \sum_{(k,l)}\left(
                1 -  2\Re\{\sum_{b\in\mathcal{B}_k}
                    \M{m}_{b,k,l}\herm\M{R}_b\M{b}_{k,l}\} + \right. \\
            \ds \left. \hspace{2.0cm}
                \left(\sum_{b\in\mathcal{B}_k}\M{m}_{b,k,l}\herm\M{R}_b\right) 
                \left(\sum_{b\in\mathcal{B}_k} \M{R}_b\herm \M{m}_{b,k,l}\right)
            \right)
    \end{array}
\end{equation}
From here, we can see that the composite pilot training matrices ($\M{R}_b$) and
the intended signal part ($\Re\{\M{R}_b\M{b}_{k,l}\}$) for any stream $(k,l)$
can be separately estimated. Assuming that we weight the intended signal part by
the corresponding $\sqrt{w_{k,l}}$, we get the modified \ac{LS} estimate for the
$l^\text{th}$ stream of \ac{UE} $k$ as
\begin{equation}
    \label{eq:ls_objective}
    \begin{array}{l}
        \ds
            \underset{\mathbf{m}_{b,k,l}}{\min.}\quad
                \frac{1}{S}
                \sum_{(k,l)}\left(1 -  2\Re\{w_{k,l}\M{u}_{k,l}\herm 
                    \sum_{b\in\mathcal{B}_k}\M{H}_{b,k}\M{m}_{b,k,l}\} +
                 \right.
            \\
        \ds
            \left. \tilde{\psi}_{k,l} +
                \sum_{b\in\mathcal{B}_k}\M{m}_{b,k,l}\herm 
                \left( 
                    \sum_{i = 1}^K\sum_{z = 1}^{L_i} 
                        \M{H}_{b,i}\herm\M{u}_{i,z}w_{i,z}
                        \M{u}_{i,z}\herm\M{H}_{b,i}
                \right)\M{m}_{b,k,l}\right)
            \text{,}
    \end{array}
\end{equation}
where $\tilde{\psi}_{k,l} \geq 0$ indicates the weighted pilot cross
interference. It is easy to see that~\eqref{eq:ls_objective} clearly corresponds
to the \ac{WMMSE} objective in~\eqref{prob:mse_min} with imperfect pilot
estimation. In fact, it is equivalent to~\eqref{prob:mse_min}, if we again let
the pilot sequence to be orthogonal and $\M{N}_b \rightarrow \M{0}$.

Finally, using~\eqref{eq:ls_objective}, we can write the transmit beamformer
design problem as
\begin{equation}
    \label{prob:mmse_downlink}
    \begin{array}{rl} 
        \ds
        \underset{\mathbf{m}_{b,k,l}}{\min.} 
            & \ds
            \sum_{(k,l)}\left(
                1 -  
                2\Re\{\sum_{b\in\mathcal{B}_k}
                    \sqrt{w_{k,l}}\M{m}_{b,k,l}\herm\M{R}_b\M{b}_{k,l}\} + 
                \right. 
            \\
            & \ds 
                \left.
                \left(\sum_{b\in\mathcal{B}_k}
                    \M{m}_{b,k,l}\herm\M{R}_b\right) 
                \left(\sum_{b\in\mathcal{B}_k}
                    \M{R}_b\herm \M{m}_{b,k,l}\right)
            \right)
            \\
        \mathrm{s.\ t.} 
            & \ds
                \sum_{k \in \mathcal{C}_b}^K\sum_{l = 1}^{L_k} 
                    \|\mathbf{m}_{b,k,l}\|^2 \leq P_b,\ b = 1,\ldots,B
            \text{.}
    \end{array}
\end{equation}
Problem~\eqref{prob:mmse_downlink} requires the knowledge of the received
training matrices $\M{R}_b$, training sequences $\M{b}_{k,l}$ and the weights
$w_{k,l}$.  Just like in \ac{SSE}, all of this can be gathered with carefully
designed \ac{TDD} pilots and feedback for the
weights~\cite{Komulainen-Tolli-Juntti-13}.

\subsection*{Downlink beamformer estimation}
Similarly to the uplink case, let the received composite downlink pilot training
matrix at \ac{UE} $k = 1,\ldots,K$ be given as
\begin{equation}
    \label{eq:downlink_training}
    \M{T}_{k} = \sum_{i = 1}^K \sum_{l = 1}^{L_i} 
                    \left(
                        \sum_{b \in \mathcal{B}_i}
                            \M{H}_{b,k}\M{m}_{b,i,l}
                    \right) \M{g}_{i,l} + \M{N}_k 
    \text{.}
\end{equation}
As the rate optimal receive beamformers are the \ac{MSE} minimizing receivers.
We can directly formulate the \ac{MMSE} estimators for receive beamformers
from~\eqref{eq:downlink_training} as
\begin{equation}
\label{eq:de_lmmse}
    \M{u}_{k,l} = {\left(\M{T}_k\M{T}_k\herm\right)}^{-1} 
                    \M{T}_{k}\M{g}_{k,l}\herm
    \text{.}
\end{equation} 
In the sequel, we consider the decentralized beamforming techniques for the
\ac{DE} approach. Note that \ac{MMSE} receive beamformer estimation is readily
decentralized and, thus, we can focus only to the downlink transmit beamformer
estimation. 

\subsection{Decentralized processing}
Due to the coherent transmission within the \ac{JP} clusters the \ac{WMMSE}
minimization, also with the \ac{DE}, is coupled among the cooperating \acp{BS}.
As the \acp{BS} are not able to to estimate the individual stream specific
pilots accurately, the fine grained decentralized processing techniques from
Sections~\ref{sec:decentralized_best_response} and~\ref{sec:decentralized_admm}
are not applicable as is. In the following, we provide modified versions for the
decentralized \ac{SSE} techniques such that the principal approaches provided in
Section~\ref{sec:sse} remains the same.

\subsubsection{\Ac{DE-BR}}
Similarly to the \ac{BR} design in
Section~\ref{sec:decentralized_best_response}, we begin assuming that all the
cooperating \acp{BS} have fixed and known transmission. The resulting beamformer
optimization for \ac{BS} $b$ can then be written as\footnote{We can ignore all
constant terms from the objective, as they do not affect the optimal results.}
\begin{equation}
    \label{prob:de_br_beamformers}
    \begin{array}{rl} 
        \ds
        \underset{\mathbf{m}_{b,k,l}}{\min.} 
            & \ds
            \sum_{k\in\mathcal{C}_b}\sum_{l = 1}^{L_k}\left(
                \M{f}_{b,k,l}\herm \M{f}_{b,k,l} - 
                2\Re\{\M{m}_{b,k,l}\herm\M{R}_b\M{b}_{k,l}\sqrt{{w}_{k,l}}\}\}
            \right)
            \\
        \mathrm{s.\ t.} 
            & \ds
                \sum_{k \in \mathcal{C}_b}^K\sum_{l = 1}^{L_k} 
                    \|\mathbf{m}_{b,k,l}\|^2 \leq P_b
            \text{,}
    \end{array}
\end{equation}
where
\begin{equation}
    \M{f}_{b,k,l} = \M{R}_b\herm\M{m}_{b,k,l} + 
                    \sum_{j \in \mathcal{B}_k \setminus \{b\}} 
                        \M{c}_{j,k,l}^{(n)}
\end{equation}
and the fixed terms from the cooperating \acp{BS} are given as
\begin{equation}
    \M{c}_{j,k,l}^{(n)} = 
        [\M{R}_j^{(n)}]\herm\M{m}_{j,k,l}^{(n)} \in \mathbb{C}^S
    \text{.}
\end{equation}
After each iteration $n$ the fixed terms are signaled within the \ac{JP}
clusters and beamformers are updated with a sufficiently small step-size
$\alpha$ as
\begin{equation}
    \label{eq:de_beamformer_update}
        \M{m}^{(n+1)}_{b,k,l} = \M{m}^{(n)}_{b,k,l} + \alpha\left(
            \M{m}^*_{b,k,l} - \M{m}^{(n)}_{b,k,l}
        \right)\ \forall\ (b,k,l)
    \text{,}
\end{equation}
where $\M{m}^*_{b,k,l}$ is the optimal solution
for~\eqref{prob:de_br_beamformers}. Keep in mind that, for \ac{DE}, the
convergence cannot be guaranteed because of the pilot estimation noise.

The beamformers can be written in a closed form expressions by evaluating the
first-order optimality conditions as
\begin{equation}
    \label{eq:de_br_beamformers}
    \M{m}_{b,k,l} = {\left(\M{R}_b\M{R}_b\herm + \I\nu_b\right)}^{-1}
        \M{R}_b\left(
            \M{b}_{k,l}\sqrt{w}_{k,l} - [\bar{\M{c}}_{b,k,l}^{(n)}]\herm
        \right)
    \text{,}
\end{equation}
where
\begin{equation}
    \bar{\M{c}}_{b,k,l}^{(n)} = 
        \sum_{j \in \mathcal{B}_k \setminus \{b\}}\M{c}_{j,k,l}^{(n)}
    \text{.}
\end{equation}
The beamformers are solved from~\eqref{eq:de_br_beamformers} by using the
bisection research for such $\nu_b$ that the power constraints are satisfied.
The basic structure of the \ac{DE-BR} algorithm is summarized in
Algorithm~\eqref{alg:de_br_decentralized}.

\subsubsection*{Signaling requirements}
The signaling requirements are apparent from~\eqref{eq:de_br_beamformers}. Each
\ac{BS} $b$ requires the knowledge of $\M{c}^{(n)}_{j,k,l}$ from the cooperating
\acp{BS} $j \in \mathcal{B}_k$ for each stream $(k,l)$. Note that vector
$\M{c}^{(n)}_{j,k,l}$ has length $S$ and, thus, the signaling overhead is also a
trade-off between the pilot resource allocation and performance.

\begin{algorithm}
    \caption{Decentralized \ac{DE-BR} algorithm for \ac{WSRMax}.}
\label{alg:de_br_decentralized}

    \begin{algorithmic}[1]
        \STATE{Initialize feasible $\M{m}_{b,k,l}\ \forall\ (b,k,l)$ and $n =
                1$.}
        \REPEAT{}
        \STATE{{\it UE\/}: Generate the \ac{MMSE} receivers $\M{u}_{k,l}\
            \forall\ (k,l)$ from~\eqref{eq:de_lmmse}.}
        \STATE{{\it UE\/}: Compute the \ac{MSE} $\epsilon_{k,l}^{(n)}\ \forall\
            (k,l)$ from~\eqref{eq:mse}.}
        \STATE{{\it UE\/}: Set the weights $w_{k,l}^{(n)}\ \forall\ (k,l)$
            from~\eqref{eq:mse_weights}.}
        \STATE{{\it BS\/}: Solve the precoders $\M{m}_{b,k,l}\ \forall\ (b,k,l)$
            from~\eqref{eq:de_br_beamformers}.}
        \STATE{{\it BS\/}: Update the next iteration precoders according
            to~\eqref{eq:de_beamformer_update}.}
        \STATE{Set $n = n + 1$.}
        \UNTIL{Desired level of convergence has been reached.}
    \end{algorithmic}
\end{algorithm}

\subsubsection{\Ac{DE-ADMM}}

Just as with the \ac{DE-BR} method, the \ac{DE-ADMM} approach follows similar
step to the \ac{SSE} \ac{ADMM} from Section~\ref{sec:decentralized_admm}.
Instead of repeating the steps in Section~\ref{sec:decentralized_admm}, we start
by formulating the estimated combined downlink signal for \ac{UE} $k$ and stream
$l$ as
\begin{equation}
\label{eq:de_admm_s}
    \bar{\M{s}}_{k,l} = \sum_{b \in \mathcal{B}_k}
                            \M{m}_{b,k,l}\herm\M{R}_b \ \forall\ (k,l)
    \text{.}
\end{equation}
Now, the \ac{ADMM} penalty term, in the $n^\text{th}$ iteration, is given as
\begin{equation}
\label{eq:de_admm_penalty}
    \Psi_{b,k,l}^{(n)} =  
        \frac{\rho}{2} \|\M{m}_{b,k,l}\herm\M{R}_b - \M{q}_{b,k,l}^{(n)}\|^2
    \text{,}
\end{equation}
where
\begin{equation}
\label{eq:de_admm_q}
    \M{q}_{b,k,l}^{(n)} = 
        \sum_{j\in\mathcal{B}_i\setminus \{b\}} 
            [\M{m}_{j,k,l}^{(n)}]\herm\M{R}_j^{(n)} - \bar{\M{s}}_{k,l} + 
            \bar{\lambda}_{k,l}^{(n)}
    \text{.}
\end{equation}
Note that, in this case, the penalty terms involve vectors of length $S$ instead
of scalars.

Using~\eqref{eq:de_admm_s} and~\eqref{eq:de_admm_penalty}, the primal
optimization problem for the beamformers $\M{m}_{b,k,l}\ \forall (b,k,l)$ and
$\bar{\M{s}}_{k,l}\ \forall\ (k,l)$ becomes 
\begin{equation}
    \label{prob:de_admm_beamformers}
    \begin{array}{rl} 
        \underset{\mathbf{m}_{b,k,l}, \M{s}_{k,l}}{\min.} &  
            \ds
            \sum_{(k,l)}
                \|\bar{\M{s}}_{k,l}\|^2 + 
                \sum_{(k,l)}\sum_{b\in\mathcal{B}_k} \Psi^{(n)}_{b,k,l} -
        \\ & \ds
            \sum_{(k,l)} \sum_{b \in \mathcal{B}_k}
                2\Re\{\sqrt{w_{k,l}}\M{b}_{k,l}\herm\M{R}_b\herm \M{m}_{b,k,l}\}
        \\ \mathrm{s.\ t.} & \ds  
                \sum_{k \in \mathcal{C}_b}^K\sum_{l = 1}^{L_k} 
                    \|\mathbf{m}_{b,k,l}\|^2 \leq P_b,\ b = 1,\ldots,B
        \text{,}
    \end{array}
\end{equation}
From~\eqref{prob:de_admm_beamformers}, we can solve the optimal
$\bar{\M{s}}_{k,l}$, for fixed $\M{m}_{b,k,l}$ as
\begin{equation}
\label{eq:de_admm_opt_s}
    \bar{\M{s}}_{k,l} = 
    \frac{\rho}{1 + \rho}
    \left(
        \sum_{b \in \mathcal{B}_k} \M{m}_{b,k,l}\herm\M{R}_{b} +
        \lambda_{k,l}^{(n)}  
    \right)
\end{equation}
Then, the dual variable update is given as
\begin{equation}
\label{eq:de_admm_dual}
    \bar{\lambda}_{k,l}^{(n+1)} = 
    \bar{\lambda}_{k,l}^{(n)} + 
    \beta\left(
        \sum_{b \in \mathcal{B}_k}\M{m}_{b,k,l}\herm\M{R}_b - 
            \bar{\M{s}}_{k,l}^{(n+1)}
    \right)
    \text{.}
\end{equation}
The dual update~\eqref{eq:de_admm_dual} differs from the \ac{SSE} \ac{ADMM} dual
update~\eqref{eq:dual_update}, by having an additional step-size parameter
$\beta$. This adds more control on how to regulate the updates. As the
intermediate beamformer updates in \ac{DE} are not exact in the sense that the
pilot-cross talk and estimation noise introduce random error into the process.
Having an more freedom to adjust the dual update, introduces more stability and
improves the performance. See Section~\ref{sec:simres} for numerical
illustration of this behavior.

For fixed~\eqref{eq:de_admm_s}, the transmit beamformer design reduces to
\begin{equation}
    \label{prob:de_admm_beamformers_avg}
    \begin{array}{rl} 
        \underset{\mathbf{m}_{b,k,l}}{\min.} &  
        \ds
            \sum_{(k,l)}\sum_{b\in\mathcal{B}_k} 
            \left( 
                \Psi^{(n)}_{b,k,l} -
                2\Re\{\sqrt{w_{k,l}}\M{b}_{k,l}\herm\M{R}_b\herm \M{m}_{b,k,l}\}
            \right)
        \\ \mathrm{s.\ t.} & \ds
            \sum_{k \in \mathcal{C}_b}^K\sum_{l = 1}^{L_k} 
                \|\mathbf{m}_{b,k,l}\|^2 \leq P_b,\ b = 1,\ldots,B
        \text{.}
    \end{array}
\end{equation}
From the first-order optimality conditions
of~\eqref{prob:de_admm_beamformers_avg}, the transmit beamformers can be written
in a closed form as
\begin{equation}
    \label{eq:de_admm_beamformers}
    \M{m}_{b,k,l} = 
        {\left(\M{R}_b\rho\M{R}_b\herm + \I\nu_b\right)}^{-1}
            \M{R}_b\left(\M{b}_{k,l}\sqrt{w}_{k,l} + \M{q}_{b,k,l} \rho\right)
    \text{.}
\end{equation}
The optimal transmit beamformers are then solved
from~\eqref{eq:de_admm_beamformers} using bisection search for the optimal
$\nu_b, b = 1,\ldots,B$ such that the power constraints are satisfied. The
\ac{DE-ADMM} algorithm is outlined in Algorithm~\ref{alg:de_admm}.

\begin{algorithm}
	\caption{\ac{DE-ADMM} algorithm for \ac{WSRMax}}
\label{alg:de_admm}
	
	\begin{algorithmic}[1]
        \STATE{{\it \ac{UE}\/}: Initialize the \ac{MMSE} receive filters
            $\M{u}_{k,l}\ \forall\ (k, l)$.}
        \STATE{{\it \ac{BS}\/}: Initialize the variables
            $\bar{\M{s}}_{k,l}^{(n)} = 0$ and dual variables
            $\bar{\lambda}_{k,l}^{(n)} = 0$ for all $(k,l)$.}
		\REPEAT{}
        \STATE{{\it \ac{BS}\/}: Update the local beamformers
            from~\eqref{eq:de_admm_beamformers}.}          
        \STATE{{\it \ac{BS}\/}: Locally update the variables $\bar{\M{s}}_{k,l},
            \M{q}_{b,k,l}$ and dual variables $\bar{\lambda}_{k,l}$
            from~\eqref{eq:de_admm_opt_s},~\eqref{eq:de_admm_q}
            and~\eqref{eq:de_admm_dual}.}
        \STATE{{\it \ac{UE}\/}: Update the receive filters $\M{u}_{k,l}\
            \forall\ (k,l)$ from~\eqref{eq:de_lmmse}.}
        \STATE{{\it UE\/}: Compute the \ac{MSE} $\epsilon_{k,l}^{(n)}\ \forall\
            (k,l)$ from~\eqref{eq:mse}.}
        \STATE{{\it UE\/}: Set the weights $w_{k,l}^{(n)}\ \forall\ (k,l)$
            from~\eqref{eq:mse_weights}.}
		\UNTIL{Desired level of convergence has been reached.}
	\end{algorithmic}
\end{algorithm}

\subsubsection*{Signaling Requirements}
The signaling requirements are equivalent to \ac{DE-BR}. In between the
beamformer updates, the cooperating \acp{BS} share the transmit beamformer
estimations $\M{m}_{b,k,l}\M{R}_b$ so that each \ac{BS} can then locally update
the $\bar{\M{s}}_{k,l}$, $\M{q}_{b,k,l}$ and $\lambda_{k,l}$.

\subsubsection{\Ac{DE-SG}}
From the objective of~\eqref{prob:mmse_downlink} it is easy to see that the
\ac{LS} estimation problem is coupled between the \acp{BS}. To come up with a
decentralized beamformer design, we resort to the stochastic gradient technique
to decouple the \ac{LS} estimation problem. To begin with, we derive the
gradient of~\eqref{prob:mmse_downlink} in terms of $\M{m}_{b,k,l}$ to be
\begin{equation}
    \label{eq:de_grad_mmse}
    \M{L}_{b,k,l} = 
        -2 \M{R}_{b} 
            \left(
                \M{b}_{k,l} \sqrt{w_{k,l}} + 
                \M{b}_{k,l} - 
                \sum_{j \in \mathcal{B}_k} \M{R}_{j}\herm\M{m}_{j,k,l}
            \right)
    \text{.}
\end{equation}
The idea in the stochastic gradient decent is, simply, to update the beamformers
in direction of the last iteration gradient. The
gradients~\eqref{eq:de_grad_mmse} are coupled. However, only the local
composites $\M{m}_{b,k,l}\herm \M{R}_{b}$ need to be shared among the
cooperating \acp{BS}. This gives us the following beamformer update routine
\begin{equation}
    \label{eq:direct_sg-update}
    \M{m}_{b,k,l}^{(n+1)} = \M{m}_{b,k,l}^{(n)} + \alpha_b \M{L}_{b,k,l} 
    \text{,}
\end{equation}
where $\M{L}_{b,k,l}$ denotes the part of~\eqref{eq:de_grad_mmse} corresponding
to \ac{BS} $b$.

Similar to Section~\ref{sec:cse_sg_dual}, the gradient
update~\eqref{eq:direct_sg-update} does not take into account the power budget.
Thus, we employ the similar dual approach to take the power budgets also into
consideration. This gives us the final beamformer update in form
\begin{equation}
    \label{eq:direct_sg-update_dual}
    \M{m}_{b,k,l}^{(n+1)} = \M{m}_{b,k,l}^{(n)} + 
                \alpha_b \M{L}_{b,k,l}  + \nu_b \M{m}_{b,k,l}
    \text{.}
\end{equation}
The outline of the \ac{SG} algorithm is given in Algorithm~\ref{alg:sg_direct}.
Note that~\eqref{eq:de_grad_mmse} is relation between the beamformer estimate
within the \ac{JP} clusters and, thus, the complete training matrices $\M{R}_b$
do not need to be available at the \acp{BS} before the backhaul signaling can
start. That is,~\eqref{eq:de_grad_mmse} can be split into training symbol level
updates
\begin{equation}
    \begin{array}{rl}
        \ds \M{L}_{b,k,l} =& \ds
        \sum_{i = 1}^S
            -2 w_{k,l} \M{R}_{b}(i)
                \sqrt{w_{k,l}} \M{b}_{k,l}(i) + 
        \\ & \ds
        2\sum_{i = 1}^S
        \left(
            \M{R}_{b}(i) 
                \sum_{j \in \mathcal{B}_k} \M{R}_{j}(i)\herm\M{m}_{j,k,l} -
            \M{b}_{k,l}(i)
        \right)
    \text{,}
    \end{array}
\end{equation}
where $\M{R}_{j}(i)$ denotes the $i^\text{th}$ column vector of $\M{R}_j$ and
$\M{b}_{k,l}(i)$ is the $i^\text{th}$ element of vector $\M{b}_{k,l}$.  This
along with the reduced computational complexity (no matrices inversion
required), can be used reduce the signaling delays even with limited
computational resources.

\subsubsection*{Signaling Requirements}
The signaling requirements are the same as with the \ac{DE-BR} and \ac{DE-ADMM}
designs with the exception that the \acp{BS} do not have to wait for the
complete feedback before starting the beamformer update routine. 

\begin{algorithm}
    \caption{\Ac{DE-SG} Ascent.}
\label{alg:sg_direct}

    \begin{algorithmic}[1]
        \STATE{Initialize feasible $\M{m}_{b,k,l}\ \forall\ (b,k,l)$ and $n =
                1$.}
        \REPEAT{}
            \STATE{{\it \ac{UE}\/}: Generate the \ac{MMSE} receivers
                $\M{u}_{k,l}\ \forall\ (k,l)$ from~\eqref{eq:lmmse}.}
            \STATE{{\it \ac{UE}\/}: Compute the \ac{MSE} $\epsilon_{k,l}^{(n)}\
                \forall\ (k,l)$ from~\eqref{eq:mse}.}
            \STATE{{\it \ac{UE}\/}: Set the weights $w_{k,l}^{(n)}\ \forall\
                (k,l)$ from~\eqref{eq:mse_weights}.}
            \STATE{{\it \ac{BS}\/}: Update the precoders $\M{m}_{b,k,l}(i)\
                \forall\ (b,k,l)$ from~\eqref{eq:direct_sg-update_dual}.}
            \STATE{{\it \ac{BS}\/}: Update the duals
                from~\eqref{eq:sg_dual_variable}}
        \UNTIL{Desired level of convergence has been reached.}
    \end{algorithmic}
\end{algorithm}


\section{Bi-directional beamformer training}
\label{sec:training}
The proposed \ac{BR}, \ac{ADMM} and \ac{SG} have similar signaling requirements.
That is, all approaches require the exchange of the effective channels along
with coherently received signals within the \ac{JP} clusters. Still, the
signaling requirements impose overhead and makes the convergence of the
algorithm slower. In order to significantly improve the rate of convergence, we
propose a bi-directional signaling scheme with multiple signaling iterations per
transmitted frame.

\begin{figure}
    \resizebox{0.95\columnwidth}{!}{%
        \begin{tikzpicture}
        \draw (0,2) -- (2,2);
        \draw (0,0) -- (2,0); 
        \draw [fill=gray!60] (2,0) rectangle (8,2);

        \draw [fill=blue!30]   (8,0)  rectangle (16,2);

        \node at (12,1) {{\Huge Data}};

        \draw [fill=darkgray] (8,0)   rectangle (8.5,2);
        \draw [fill=lightgray!60]  (8.5,0) rectangle (9.0,2);
        \draw [fill=darkgray] (9.0,0) rectangle (9.5,2);
        \draw [fill=lightgray!60]  (9.5,0) rectangle (10.0,2);

        \draw [fill=gray!60]  (14,0)  rectangle (20,2);
        
        \draw (20,2) -- (22,2);
        \draw (20,0) -- (22,0);

        \node at (1,0.90) {{\Huge $\cdots$}};
        
        \node at (5,1)  {{\Huge Frame $n\hspace{-2mm}-\hspace{-2mm}1$}};
        \node at (17,1) {{\Huge Frame $n\hspace{-2mm}+\hspace{-2mm}1$}};
        
        \node[white] at (8.30,1) {{\Huge U}};
        \node at (8.77,1) {{\Huge D}};
        \node[white] at (9.30,1) {{\Huge U}};
        \node at (9.77,1) {{\Huge D}};
        
        \node at (21,0.90) {{\Huge $\cdots$}};
        \draw [
                thick,
                decoration={brace,
                    amplitude=10pt,
                    mirror,
                    raise=0.25cm
                },
                decorate
            ] (8,0) -- (14,0);
        
        \node at (11,-1.5) {\Huge Frame $n$};
            
        \draw [
                thick,
                decoration={brace,
                    amplitude=10pt,
                    raise=0.25cm
                },
                decorate
            ] (8,2) -- (10,2);
            
        \node at (9,3.0) {\Huge Beamformer Signaling};
        \end{tikzpicture}
    }
\caption{\ac{TDD} frame structure with two bi-directional beamformer signaling
    iterations.}
\label{fig:frame_structure}
\end{figure}
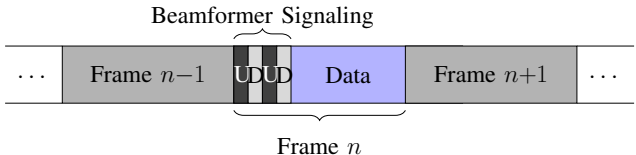

We employ bi-directional \ac{UL}/\ac{DL} signaling for \ac{TDD} systems with
similar frame structure to the training scheme proposed
in~\cite{Shi-Berry-Honig-TSP-14}. The bi-directional signaling allows direct
exchange of the effective \ac{UL} and \ac{DL} channels from the corresponding
precoded UL/DL pilot signals. The signaling sequence occupies a fraction
$\gamma$ of the \ac{DL} frame. The remaining portion ($1 - \gamma$) of the frame
is reserved for the transmitted data.  The frame structure is illustrated in
Fig.~\ref{fig:frame_structure}, where D and U denote \ac{DL} and \ac{UL} pilots,
respectively. Here, we assume that the effective channels are perfectly
estimated in each iteration.

After a signaling iteration (\ac{UL}/\ac{DL} sequence), each \ac{BS} $b =
1,\ldots,B$ has up-to-date information on the effective DL channels with the
current receivers applied ($\mathbf{u}_{k,l}\herm\mathbf{H}_{b,k}\ \forall\
(k,l)$).  Successive \ac{UL}/\ac{DL} signaling iterations allow fast beamformer
signaling and can potentially offer improved tracking for the channel changes.
Conventionally, the beamformer signaling is contained in the precoded
demodulation and channel sounding pilots, which allows only one \ac{UL}/\ac{DL}
iteration per transmitted frame~\cite{Komulainen-Tolli-Juntti-13}. This results
in significantly slower beam coordination.  Note that the weight factors can be
incorporated into the effective channels~\cite{Komulainen-Tolli-Juntti-13},
which further reduces the signaling overhead. 

\subsection*{Feedback Quantization}
The feedback signaling information has to be quantized before it is exchanged
over a feedback channel or the backhaul.  Thus, robustness to the quantization
errors is crucial for any design realizable in practice. In addition,
quantization reduces the backhaul utilization, and, in turn, enables more
elaborated iteration process. 

\begin{figure}
    \centering
    \includegraphics[width=\columnwidth]{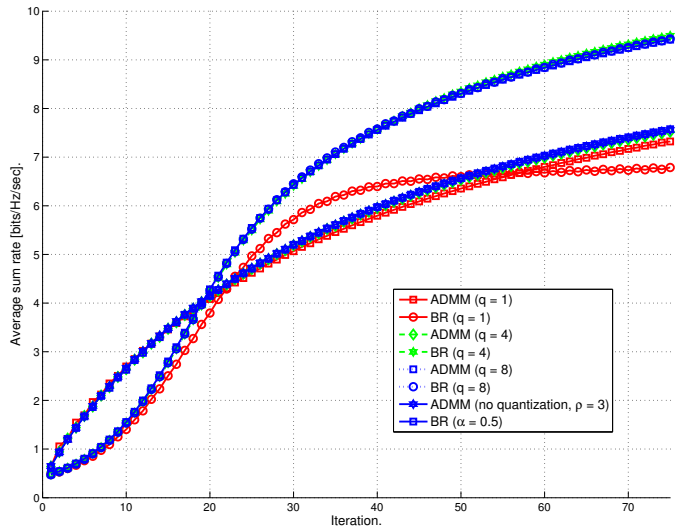}
    \caption{Differential feedback signaling with $q$-bit quantization.}
\label{fig:differential_quantization}
\end{figure}

As the beamformer training information is strongly correlated between the
iterations, the consequent signaling iterations contain large amounts of
redundant information. To exploit this correlation, we propose a differential
signaling scheme, where each \ac{BS} signals the quantized difference of the
latest and previous iteration signals. The feedback information is thus
iteratively improved as the algorithm progresses. Furthermore, a smoothing
operator can be added to improve the convergence properties, that is, a feedback
symbol $s$ is updated as
\begin{equation}
    s^{i+1} = s^i + \beta (s - s^i)
    \text{,}
\end{equation}
where $\beta$ denotes the update step size.  The I and Q branches of the complex
feedback symbols are quantized separately.
Fig.~\ref{fig:differential_quantization} demonstrates the convergence behavior
of the \ac{BR} and \ac{ADMM} methods for varying levels of quantization (see
Section~\ref{sec:simres} for more details on the simulation environment). Both
approaches are clearly capable of coping with the quantized feedback even with
small quantization levels. It is also evident that the \ac{ADMM} approach
provides better initial rate of convergence, while the \ac{BR} design speeds up
after a few initial iterations. 


\section{User Admission}
\label{sec:useradm}

Overloaded initialization, in the sense that there are more active spatial data
streams than available \ac{DoF}, has been proposed in various publications as an
efficient user admission design~\cite{Shi-Razaviyayn-Luo-He-11}. As a result of
the transceiver iteration, the excess streams will be dropped, i.e., the
corresponding beamformers will get zero power~\cite{Shi-Razaviyayn-Luo-He-11}.

For static channels, the overloaded initialization can be used as a low
complexity user allocation approach, particularly, for complex systems with a
large number of users. This is not particularly convenient for time correlated
channel models, where the channel conditions change in time. In such cases, it
is more beneficial to dynamically change the user allocation to better reflect
the changing channel conditions. However, reintroducing the dropped users is
difficult as the priority weight factors of the reintroduced users should be
proportional to the active users.  Furthermore, once the spatial compatibility
of the active streams is close to a local optima, it is difficult to reintroduce
a stream to the system in such a way that the reintroduced streams potentially
improve performance of the existing setup. In this case, it is likely that the
reintroduced streams will be dropped, due to the spatial incompatibility.

To overcome the degraded beamformer compatibility in time correlated channels,
we propose beamformer reinitialization after a given number of iterations. This
effectively performs periodic user selection. The reinitialization has a
significant impact on the system performance and has been numerically evaluated
in Section~\ref{sec:simres}. 

\subsection{Varying beamformer signaling length}

We can exploit the fact that the performance loss is caused by insufficient
beamformer convergence. A straightforward approach is to make the beamformer
signaling part of each frame longer. This gives more time for the beamformers to
converge. However, this also increases the signaling overhead and, which may
become excessive for the later iterations as the beamformers have already
sufficiently converged and user selection has occurred. 

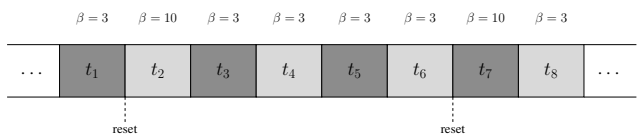
\begin{figure}[h]
    \resizebox{0.95\columnwidth}{!}{%
        \begin{tikzpicture}
        \draw (0,2) -- (2,2);
        \draw (0,0) -- (2,0); 
        \node at (1,0.90) {{\huge $\cdots$}};
        
        \foreach \x [count=\i] in {2,7,...,17} {
            \draw [fill=darkgray!60] (\x,0) rectangle (\x+2.5,2);
            \edef\c{\i};
            \pgfmathparse{int(\c*2-1)};
            \xdef\c{\pgfmathresult};
            \node at (\x+1.25, 1) {{\huge $t_{\c}$}};
            \draw [fill=lightgray!60]  (\x+2.5,0) rectangle (\x+5,2);
            \pgfmathparse{int(\c+1)};
            \xdef\c{\pgfmathresult};
            \node at (\x+2.5+1.25, 1) {{\huge $t_{\c}$}};
        }
        
        \foreach \x [count=\i] in {2,4.5,...,19.5} {
            \ifthenelse{\i=2 \OR \i=7}{
                \node at (\x+1.25,3) {{\Large $\beta=10$}};
            }{
            \node at (\x+1.25,3) {{\Large $\beta=3$}};
        };
    }
    
    \draw (22,2) -- (24,2);
    \draw (22,0) -- (24,0);
    \node at (23,0.90) {{\huge $\cdots$}};
    
    \draw[dashed] (4.5,0) -- (4.5,-1);
    \node at (4.5,-1.25) {{\Large reset}};
    
    \draw[dashed] (17.0,0) -- (17.0,-1);
    \node at (17.0,-1.25) {{\Large reset}};
    
    \end{tikzpicture}
}
\caption{An example of varying number beamformer signaling iterations with
    respect to the user reinitialization index.}
\label{fig:dyn_bit}
\end{figure}

We propose a varying length beamformer signaling among the frames, where the
beamformer signaling interval is longer after each reinitialization point and
shorter for the subsequent frames. This improves the inherent trade-off between
the signaling overhead and beamformer convergence. This scheme has been
illustrated in Fig.~\ref{fig:dyn_bit}, where the number of signaling iterations
is fixed to $\beta = 10$ after each reinitialization frame and to $\beta = 3$
for all the other frames. Varying the signaling lengths allows the algorithm to
achieve most of the performance during the first frame while having most of the
performance penalty in duration of one frame. This penalty is compensated during
the subsequent frames with less beamformer signaling iterations.  

%
%
%
%
%


\subsection{Delayed beamformer indexing}

Having a varying length beamformer training iterations depending on the frame
index may require excessive planning in smaller (femto sized) systems as the
\ac{TDD} frame structure has to be globally identical in order to assure limited
pilot signal contamination by the interfering transmissions. To this end, we
propose a more flexible alternative method to improve the diminished system
performance after each beamformer reinitialization.

For reasonably slow fading channels, we may assume that the changes in the
channels between two consecutive frames is not overly drastic. Thus, the
performance decrease for fixed beamformers between two frames is only minor.
This assumption may be exploited with the beamformer reinitialization by
delaying the beamformer indexing in the sense that, as the trained
transmit/receive beamformers are reinitialized, the beamformers before the
reinitialization are used for the actual data transmission until the trained
beamformers have converged to sufficiently high performance. 

Note that the receive beamformers can be always assumed to be up-to-date as the
active transmit beamformers can be estimated directly from the demodulation
pilots (see Fig.~\ref{fig:frame_structure}). Here, $\M{m}(t_i)$ denotes the
active beamformers generated in the frame $t_i$. By delaying the beamformers for
one iteration after the reset, the degradation in the achievable rate is
significantly reduced. On the second frame after the reinitialization, the
active beamformers have already converged to overcome most of the negative
impact from the beamformer reset and can be switched as the active beamformers
for the actual data transmission.
\begin{figure}[h]
    \resizebox{0.95\columnwidth}{!}{%
        \begin{tikzpicture}
        \draw (0,2) -- (2,2);
        \draw (0,0) -- (2,0); 
        \node at (1,0.90) {{\huge $\cdots$}};
        
        \foreach \x [count=\i] in {2,7,...,17} {
            \draw [fill=darkgray!60] (\x,0) rectangle (\x+2.5,2);
            \edef\c{\i};
            \pgfmathparse{int(\c*2-1)};
            \xdef\c{\pgfmathresult};
            \node at (\x+1.25, 1) {{\huge $t_{\c}$}};
            \draw [fill=lightgray!60]  (\x+2.5,0) rectangle (\x+5,2);
            \pgfmathparse{int(\c+1)};
            \xdef\c{\pgfmathresult};
            \node at (\x+2.5+1.25, 1) {{\huge $t_{\c}$}};
        }
        
        \foreach \x [count=\i] in {2,4.5,...,19.5} {
            \ifthenelse{\i=2 \OR \i=7}{
                \edef\c{\i};
                \pgfmathparse{int(\c-1)};
                \xdef\c{\pgfmathresult};
                \node at (\x+1.25,3) {{\Large $\mathbf{m}(t_{\c})$}};
            }{
            \node at (\x+1.25,3) {{\Large $\mathbf{m}(t_{\i})$}};
        };
    }
    
    \draw (22,2) -- (24,2);
    \draw (22,0) -- (24,0);
    \node at (23,0.90) {{\huge $\cdots$}};
    
    \draw[dashed] (4.5,0) -- (4.5,-1);
    \node at (4.5,-1.25) {{\Large reset}};
    
    \draw[dashed] (17.0,0) -- (17.0,-1);
    \node at (17.0,-1.25) {{\Large reset}};
    
    \end{tikzpicture}
}
\caption{An illustration of delayed transmit beamformer indexing.}
\label{fig:delayed_precoders}
\end{figure}
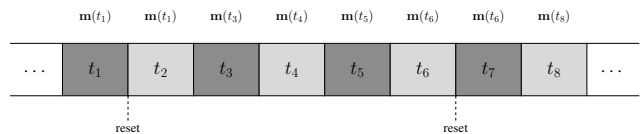

%
%
%
%

In the end, this technique utilizes two sets of beamformers. First set consists
of the beamformers that are being trained and iteratively exchanged among the
interfering transmitters using the bi-directional signaling portion of the frame
structure. The second set of beamformers are the ones that used in the current
frame to actually transmit the data. 



\section{Numerical Examples}
\label{sec:simres}

\begin{figure}[ht]
    \centering
    \includegraphics[width=\columnwidth]{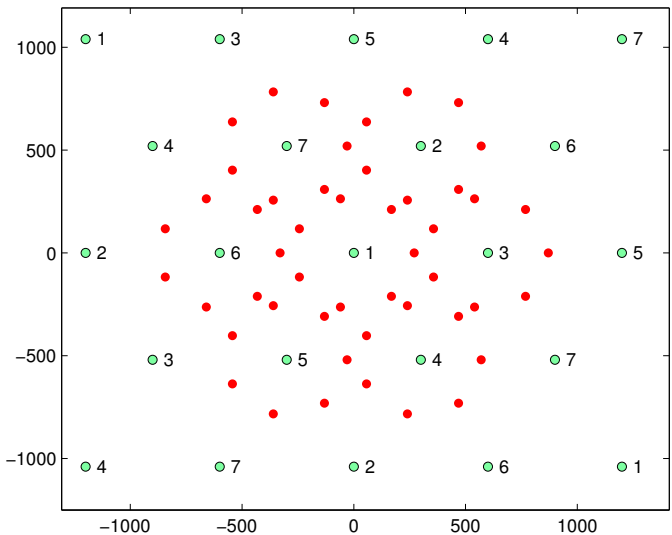}
    \caption{Base station and cell edge user terminal deployment in 7-cell wrap
        around model with $K_b = 7$ in each cell.}
\label{fig:wrap7-env}
\end{figure}

The simulations are carried out using a 7-cell wrap around model, where the
distance between the \acp{BS} is $600\text{m}$. The path loss exponent for the
user terminals is fixed to $3$. The number of transmit and receive antennas are
set to $N_\text{T} = 4$ and $N_\text{R} = 2$, respectively. There are $K_b = 7$
user terminals that are evenly distributed on the cell edge around each \ac{BS}.
In total, there are $K = BK_b = 49$ users in the network. We assume full
cooperation, i.e., all users are coherently served by every BS in the system. In
practice, practical constraints such as pilot contamination will limit the
number of active users per-\ac{BS}. The number of active spatial stream per
users is limited to one. The simulation environment is illustrated in
Fig.~\ref{fig:wrap7-env}.

The \ac{SNR} is defined on the cell edge from the closest \ac{BS} $b$, i.e,
$\text{SNR} = \frac{g_{b,k}P_b}{\sigma_k^2}$, where $g_{b,k}$ denotes the
corresponding path loss. The channels are generated with Jakes' Doppler spectrum
model. The channel coherence time is defined by normalized user terminal
velocity $t_\text{S}f_\text{D}$, where $t_\text{S}$ and $f_\text{D}$ are the
backhaul signaling rate and the maximum Doppler shift, respectively. Simulations
are performed for two user velocity scenarios $t_\text{S}f_\text{D} = 0.01$ and
$t_\text{S}f_\text{D} = 0.025$ that correspond to user velocities of $2.7$ km/h
and $6.9$ km/h, respectively.  The block fading model assumes that the channels
remain constant during the transmission of each frame, and the changes occur
in-between the frames. If not mentioned otherwise, the \ac{ADMM} simulations are
done with $\rho = 3$ and \ac{BR} simulations are performed with $\alpha = 0.5$.
Summary of the simulation parameters is listed in Table~\ref{tbl:simparams}.

\begin{table} 
    \caption{Simulation parameters.}
\label{tbl:simparams}
    \begin{tabular}{cc}
        \toprule 
        \bfseries Parameter & \bfseries Value \\ 
        \midrule \midrule 
        Number of \acp{UE} ($K$)  & $49$ \\
        \midrule 
        Number of cells ($B$)  & $7$ \\
        \midrule 
        Number of \acp{UE} per cell & $7$ \\
        \midrule 
        \ac{BS} antennas ($N_\text{T}$)  & $2$ \\
        \midrule 
        \ac{UE} antennas ($N_\text{R}$)  & $2$ \\
        \midrule 
        Distance between adjacent \acp{BS} & $600$ m \\
        \midrule 
        The path loss exponent & $3$ \\
        \midrule 
        Signaling rate ($t_{S}$)  & $2$ ms \\
        \midrule 
        Carrier frequency & $2$ GHz \\
        \midrule 
        \ac{UE} velocities  & $0$ km/h, $2.7$ km/h and $6.9$ km/h \\
        \bottomrule 
    \end{tabular} 
\end{table}

The bi-directional signaling overhead is considered using coefficient $\gamma
\in [0,1]$, so that the achievable rate is defined as $(1 - \gamma) R$.  The
overhead coefficient $\gamma$ defines the fraction of the frame length, which is
reserved for the signaling sequence.  The number of \ac{UL}/\ac{DL} signaling
iterations is denoted by $\text{BIT}$ (bi-directional iterations).  By this
notation, the complete frame length is $2\gamma^{-1}\text{BIT}$. We assume that
the \ac{UE} feedback channels are slow in the sense that the stream specific
weights $w_{k,l}\ \forall\ (k,l)$ can be exchanged only once per frame. That is,
the bidirectional iteration, within a frame, only involves \ac{TDD} based
beamformer signaling.

\subsection{Stream Specific Estimation Methods}

\begin{figure}
    \centering 
    \includegraphics[width=\columnwidth]{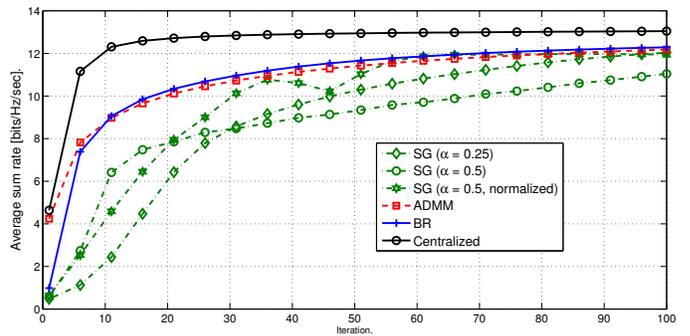} 
        \caption{Average achievable sum rate per \ac{BS} in static channel.}
\label{fig:cse_method_comparison}
\end{figure}

The proposed \ac{SSE} methods from Section~\ref{sec:sse} are compared in
Fig.~\ref{fig:cse_method_comparison}. The asymptotic performance of all of the
proposed designs are comparable and the differences in performance are mostly
related to the rate of convergence. It can be seen that the \ac{ADMM} design
provides the fastest initial convergence. However, the performance becomes
comparable to the \ac{BR} method after few initial iterations. The \ac{SG}
approach has slower rate of convergence. However, the step size normalization
does help.  When taking into account the lower complexity, the \ac{SG} approach
can be seen to be a viable alternative to the more complex decentralized
methods.

\begin{figure}
    \centering 
    \includegraphics[width=\columnwidth]{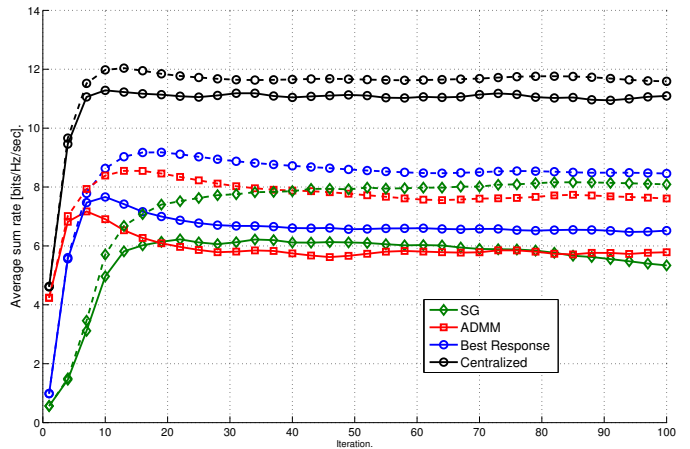} 
    \caption{Average achievable sum rate per BS in time correlated channels with
        \ac{UE} speeds $2.7$ km/h (dashed) and $6.9$ km/h (solid).}
\label{fig:time_correlated_cse_comparison}
\end{figure}

In Fig.~\ref{fig:time_correlated_cse_comparison}, the \ac{SSE} methods are
compared using time correlated channels. The dashed lines show the performance
for \ac{UE} speed of $2.7$ km/h and the solid lines show the performance with
\ac{UE} speed of $6.9$ km/h. The time correlated indicates similar behavior to
the static channel. The rate of convergence of the \ac{ADMM} method is faster in
the beginning, which results in good performance for the first, few iterations.
However, the reduced rate of convergence for the later iterations, results in
somewhat diminished capability to follow the channel changes. The \ac{BR} design
performs the best in the later phase. On the other hand, the \ac{SG} based
beamforming provides competitive performance, considering the greatly reduced
computational complexity.

\subsection{Direct Estimation}
\begin{figure}
    \centering
    \includegraphics[width=\columnwidth]{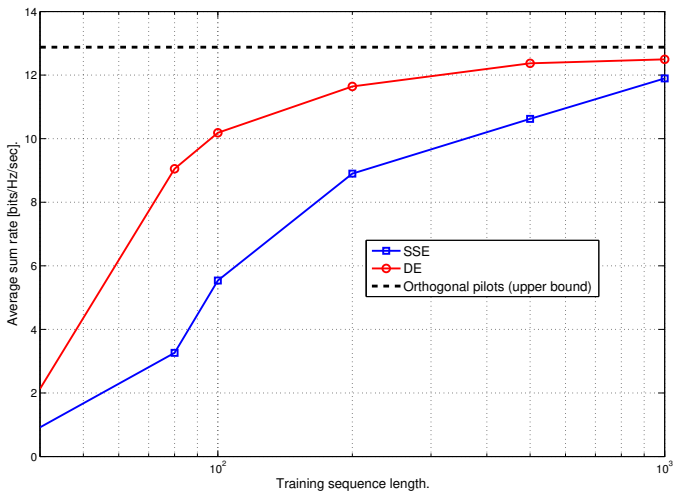}
    \caption{Centralized \ac{DE} behavior for varying training sequence lengths
    in constant channel.}
\label{fig:de_training_length}
\end{figure}

Fig.~\ref{fig:de_training_length} demonstrates the performance of the
centralized \ac{DE} and \ac{SSE} as the length of the pilot training sequence is
varied. Here, the \ac{SSE} beamformer design is done with the same pilots as the
\ac{DE}, only ignoring the pilot cross-talk and estimation error. It is easy to
confirm that \ac{DE} has clear advantage, when the pilot contamination levels
are high. On the other hand, it should be noted that with sufficiently long
pilot sequences the pilots can be made fully orthogonal, which reduces the
performance gap with large pilot lengths.

\begin{figure}
    \centering
    \includegraphics[width=\columnwidth]{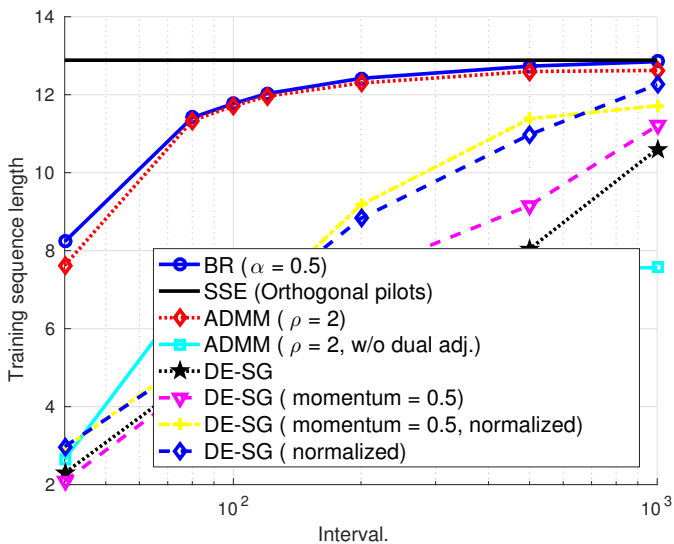}
    \caption{Comparison of the decentralized \ac{DE} methods with varying
             training sequence lengths in constant channel.}
\label{fig:decentralized_de}
\end{figure}

The impact of the pilot sequence length on the decentralized \ac{DE} processing
is shown in Fig.\ref{fig:decentralized_de}. It \ac{DE-ADMM} and \ac{DE-BR}
methods have clearly comparable performance. While the \ac{DE-SG} design
requires larger pilot lengths, to achieve comparable asymptotic performance. For
the \ac{DE-SG}, momentum and step-size normalization can be seen to
significantly improve the performance.

\begin{figure}
    \centering
    \includegraphics[width=\columnwidth]{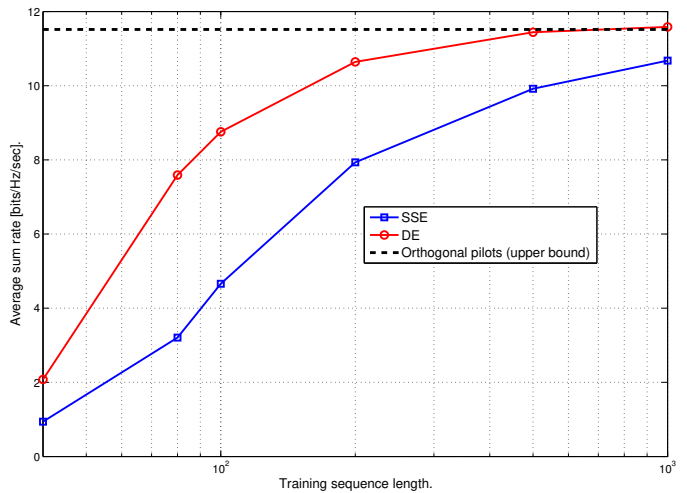}
    \caption{\ac{DE} behavior for varying training sequence lengths in time
        correlated channels with \ac{UE} speed $2.7$ km/h.}
\label{fig:de_centralized_tc_slow}
\end{figure}

Figs.~\ref{fig:de_centralized_tc_slow} and~\ref{fig:de_centralized_tc_slow}
show the performance of the centralized \ac{DE} in time correlated channel with
\ac{UE} speeds $2.7$km/h and $6.9$km/h, respectively. The time correlated
behaviour is similar to the constant channel performance. As the \ac{UE} speed
grows, the gap between the \ac{SSE} and \ac{DE} methods diminishes. This is due
the fact that both methods have similarly out-of-date \ac{CSI} and beamforming
gain is no longer available.

\begin{figure}
    \centering
    \includegraphics[width=\columnwidth]{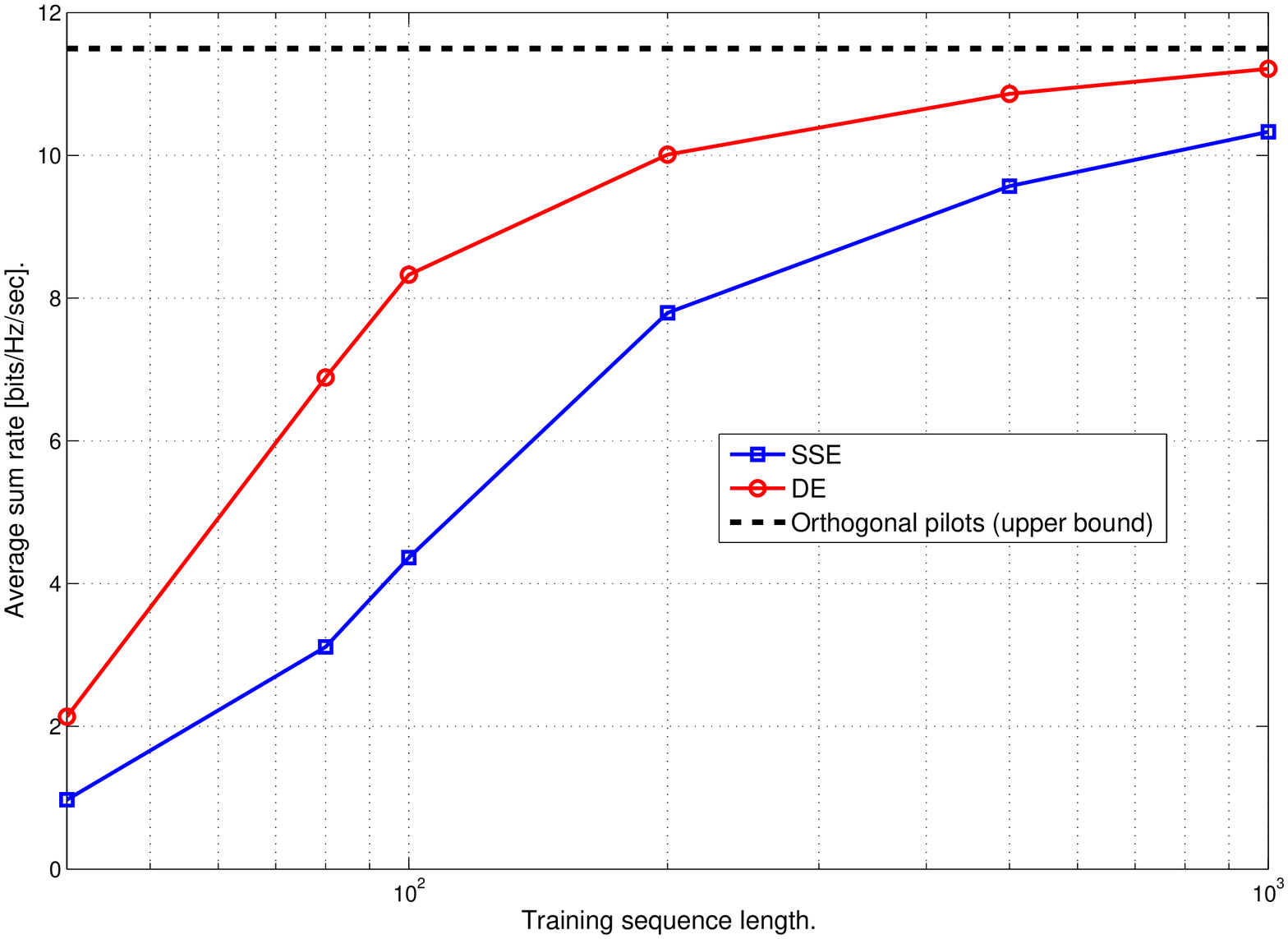}
    \caption{\ac{DE-BR} behavior for varying training sequence lengths in time
        correlated channels with \ac{UE} speed $6.9$ km/h.}
\label{fig:de_br_tc_fast}
\end{figure}

\subsubsection{\ac{DE-BR}}
\begin{figure}
    \centering
    \includegraphics[width=\columnwidth]{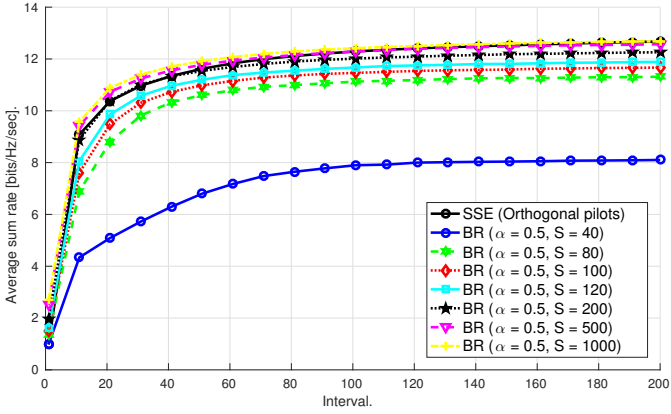}
    \caption{\ac{DE-BR} performance for varying training sequence lengths in
        constant channel.}
\label{fig:de_br}
\end{figure}

From Fig.~\ref{fig:de_br}, it is evident that the \ac{BR} will convergence to
match the \ac{SSE} performance, given long enough pilot sequences. Also, the
significance of the pilot sequence length diminishes when the pilot sequence
length grows larger than the number of interfering streams in the system, i.e.,
when $S$ grows larger than $49$. 

\begin{figure}
    \centering
    \includegraphics[width=\columnwidth]{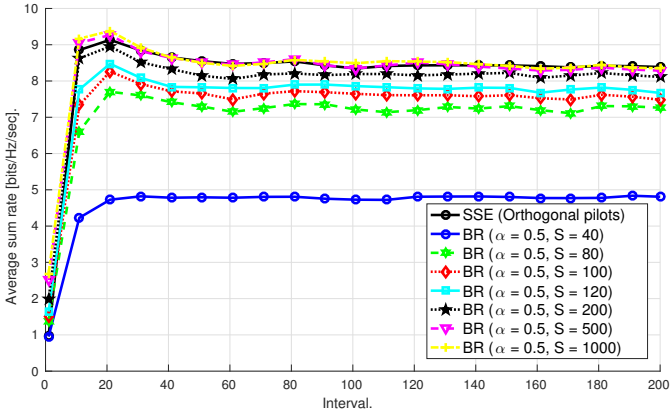}
    \caption{\ac{DE-BR} behavior for varying training sequence lengths in time
        correlated channels with \ac{UE} speed $2.7$ km/h.}
\label{fig:de_br_tc_slow}
\end{figure}

The behavior of the \ac{DE-BR} in time correlated channels are shown in
Figs.~\ref{fig:de_br_tc_slow} and~\ref{fig:de_br_tc_fast}. Here, the orthogonal
pilot allocation upper bound is generated by using the \ac{BR} method
from~\ref{sec:decentralized_best_response}. Again, we can see that as $S$ grows
larger that there are interfering stream, the difference in performance is
neglectible. 

\begin{figure}
    \centering
    \includegraphics[width=\columnwidth]{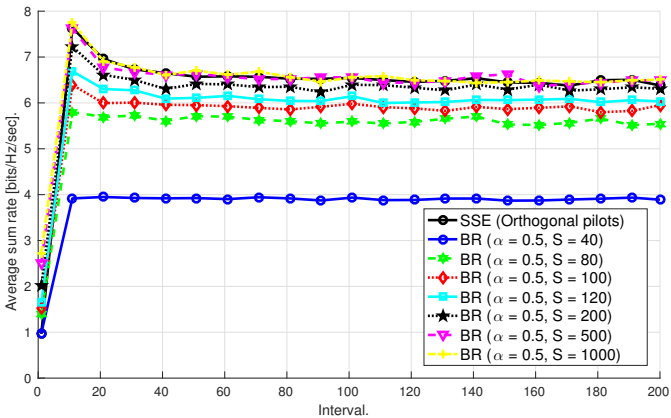}
    \caption{\ac{DE-BR} behavior for varying training sequence lengths in time
        correlated channels with \ac{UE} speed $6.9$ km/h.}
\label{fig:de_centralized_tc_fast}
\end{figure}

\subsubsection{\ac{DE-ADMM}}
\begin{figure}
    \centering
    \includegraphics[width=\columnwidth]{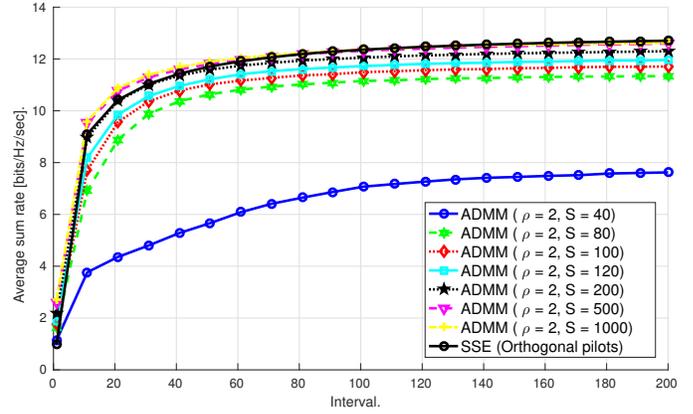}
    \caption{\ac{DE-ADMM} performance for varying training sequence lengths in
        constant channel.}
\label{fig:de_admm}
\end{figure}

\begin{figure}
    \centering
    \includegraphics[width=\columnwidth]{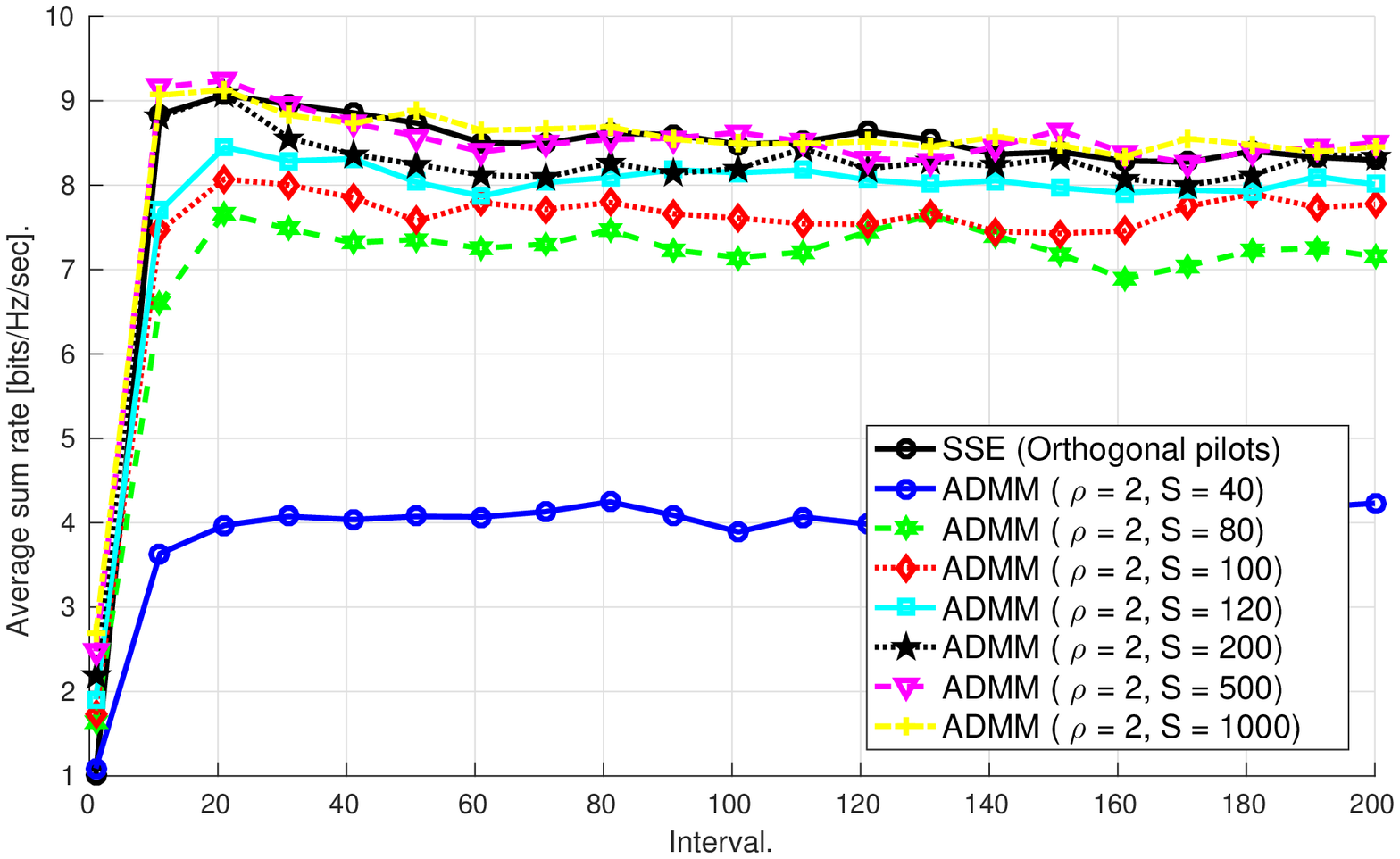}
    \caption{\ac{DE-ADMM} behavior for varying training sequence lengths in time
        correlated channels with \ac{UE} speed $2.7$ km/h.}
\label{fig:de_admm_tc_slow}
\end{figure}

\begin{figure}
    \centering
    \includegraphics[width=\columnwidth]{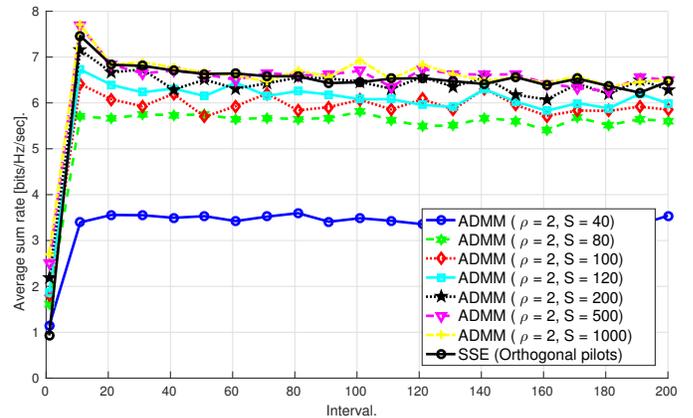}
    \caption{\ac{DE-ADMM} behavior for varying training sequence lengths in time
        correlated channels with \ac{UE} speed $6.9$ km/h.}
\label{fig:de_admm_tc_fast}
\end{figure}

When not otherwise stated the dual updates~\ref{eq:de_admm_dual} are done using
$\beta = 1/\rho$. Fig.~\ref{fig:de_admm} demonstrates the \ac{DE-ADMM} method
convergence behavior. Performance in time correlated channels is shown in
Figs.~\ref{fig:de_admm_tc_slow} and~\ref{fig:de_admm_tc_fast}. The performance
can be seen to be nearly identical to the \ac{DE-ADMM} approach. 

\subsubsection{\ac{DE-SG}}
\begin{figure}
    \centering
    \includegraphics[width=\columnwidth]{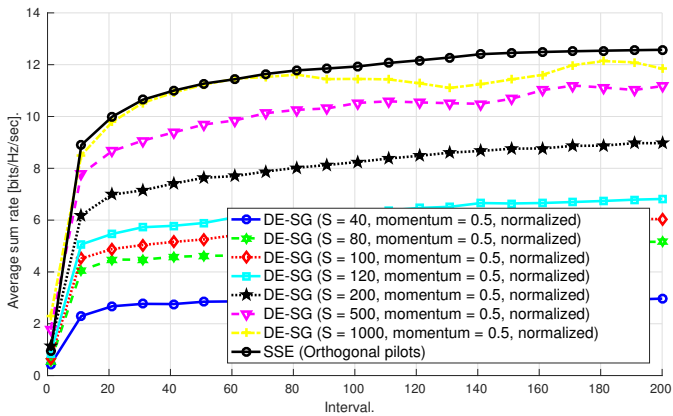}
    \caption{\ac{DE-SG} performance for varying training sequence lengths in
        constant channel.}
\label{fig:de_sg}
\end{figure}

Fig.~\ref{fig:de_sg} shows the performance of the decentralized \ac{DE-SG}
method.  The \ac{DE-SG} design can be seen to approach the performance of
\ac{SSE} design without pilot contamination as the training sequence length
becomes sufficiently large. 

\begin{figure}
    \centering
    \includegraphics[width=\columnwidth]{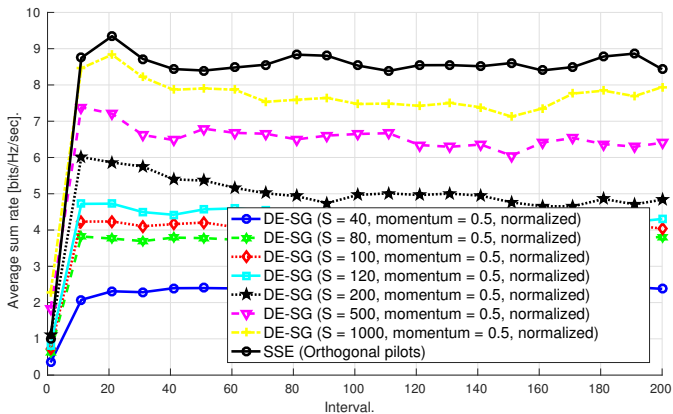}
    \caption{\ac{DE-SG} behavior for varying training sequence lengths in time
        correlated channels with \ac{UE} speed $2.7$ km/h.}
\label{fig:de_sg_tc_slow}
\end{figure}

\begin{figure}
    \centering
    \includegraphics[width=\columnwidth]{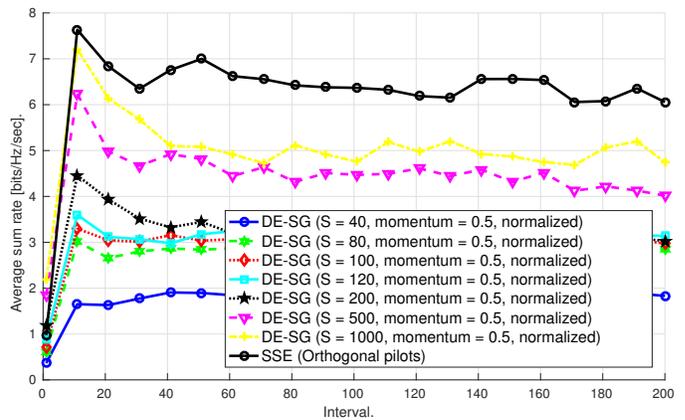}
    \caption{\ac{DE-SG} behavior for varying training sequence lengths in time
        correlated channels with \ac{UE} speed $6.9$ km/h.}
\label{fig:de_sg_tc_fast}
\end{figure}

\ac{DE-SG} performance in time correlated channels is show
in Fig.~\ref{fig:de_sg_tc_slow} and Fig.~\ref{fig:de_sg_tc_fast}. In comparison to
\ac{DE-BR} and \ac{DE-ADMM}, we can see that the \ac{SG} is more sensitive to
the pilot sequence length.  

\subsection{User Admission}

For the user admission, we lower the number of transmit antennas to $N_\text{T}
= 2$. This makes the system overloaded in the sense that there are more
initialized beamformers than there are available \ac{DoF}. There are $K = 49$
initialized streams, while there are only $B N_\text{T} = 28$
degrees-of-freedom. As discussed in Section~\ref{sec:useradm}, the excess
streams get dropped during the beamformer iteration, which effectively means
that user selection is performed. The performance of the user admission methods
is evaluated with the \ac{BR} design. 

\begin{figure}
    \centering
    \includegraphics[width=\columnwidth]{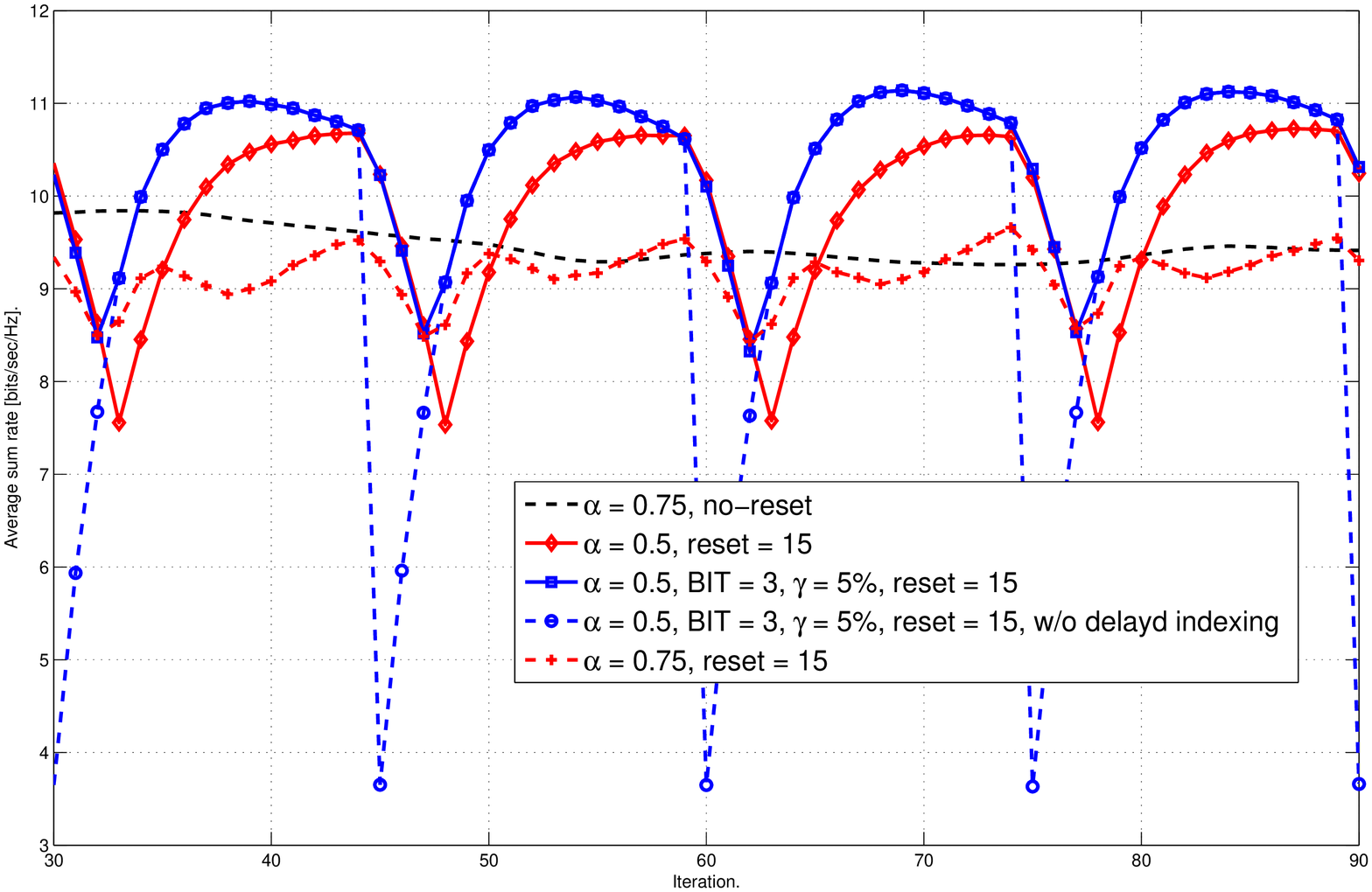}
    \caption{Average system performance using periodic reset with \ac{UE} speed
        $2.7$ km/h and \ac{SNR} $= 20$dB.}
\label{fig:usradm-slow}
\end{figure}

\begin{figure}
    \centering
    \includegraphics[width=\columnwidth]{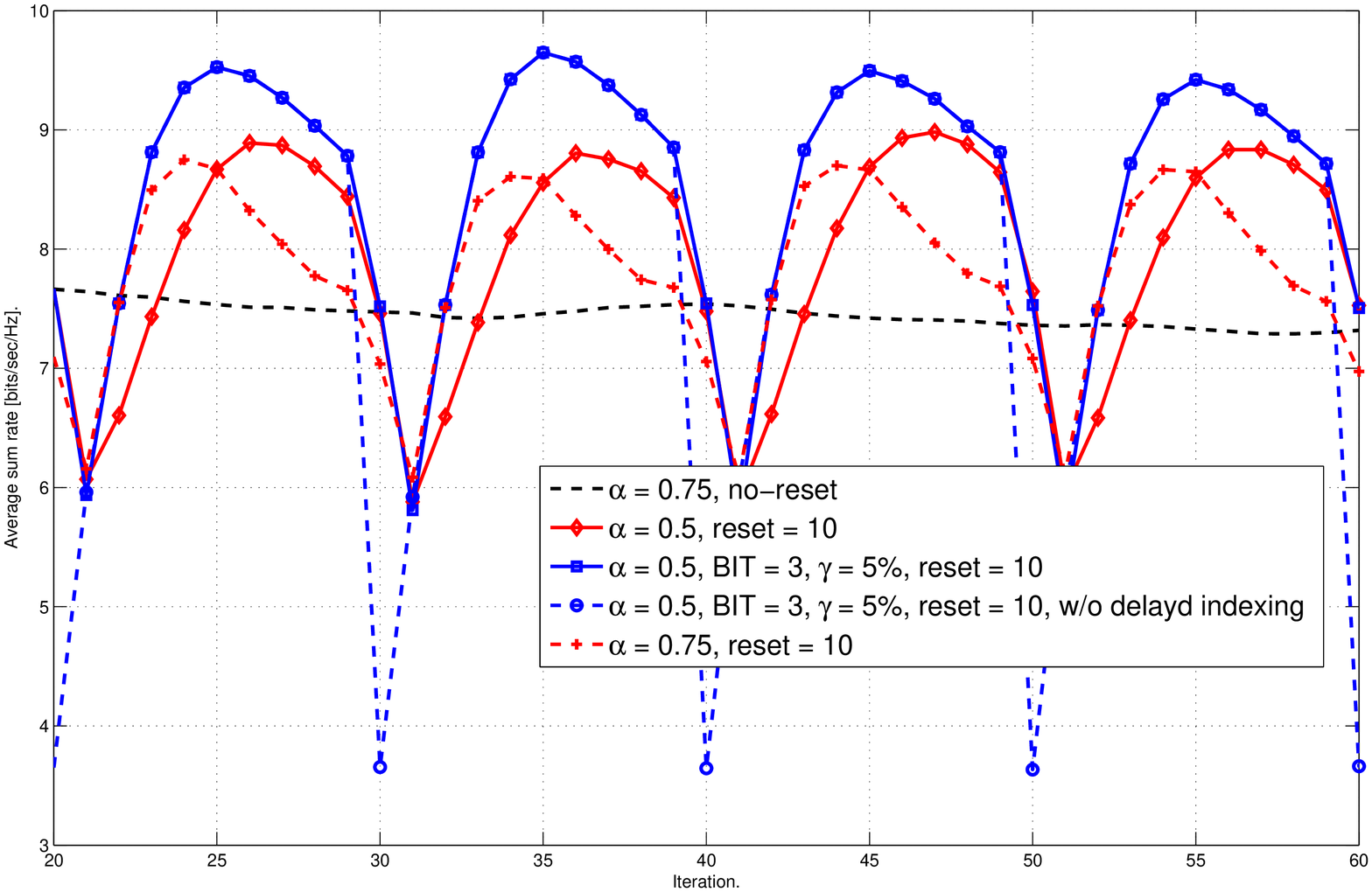}
    \caption{Average system performance using periodic reset with \ac{UE} speed
        $6.9$ km/h and \ac{SNR} $= 20$dB.}
\label{fig:usradm-fast}
\end{figure}

In Fig.~\ref{fig:usradm-slow} and~\ref{fig:usradm-fast}, the system performance
is shown in time correlated channels.  Before each reinitialization, the
beamformers are stored for the delayed indexing. Clearly, as the channels
change, the initial user selection becomes inefficient and periodically
initializing the beamformers allows the user allocation to better adjust to the
changing channel conditions. The delayed indexing significantly improves the
performance while retraining the beamformers. The bi-directional signaling can
be seen to improve the stability of the algorithm behavior as well as the
convergence properties. As the channel changes more aggressively the performance
of the delayed indexing beamformers is diminished.

\begin{figure}[h]
    \includegraphics[width=\columnwidth]
        {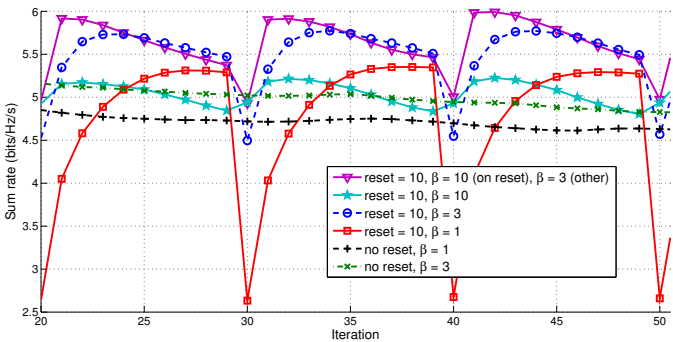}
    \caption{Average sum rate behavior of the varying length beamformer
        signaling  with 10 frame reset interval, \ac{UE} speed $6.9 km/h$ and
        $\gamma = 0.02$.}
\label{fig:simres_varying}
\end{figure}

From Fig.~\ref{fig:simres_varying}, we can observe two types of benefits from
the varying length signaling iterations.  First, the performance degradation
after the beamformer reinitialization is reduced. Secondly, the performance of
the following iterations is improved. This is due to the benefit of letting the
beamformers convergence for 10 iterations during the first frame after the
reinitialization, which results in higher improved spatial compatibility between
the transmissions on the following iterations and leads to improved system
performance.

\section{Conclusions}
\label{sec:conclusions}

We have proposed decentralized transceiver designs for coherent CoMP WSRMax. We
considered orthogonal pilot resource allocation without pilot estimation noise
and scenarios with non-orthogonal and noise pilots. Along with low complexity
and signaling overhead transceiver designs, we provided novel techniques for
user admission and beamformer training.  Numerical results indicated that our
designs provide good performance and stability even with time correlated channel
conditions.

\appendices

\section{Simplification of~\eqref{eq:summse_admm_lag_0}}
\label{ap:admm_reformulation}

To simplify expression~\eqref{eq:summse_admm_lag_0}, we can eliminate the
auxiliary variables $s_{k,l,b,i,z}\ \forall\
(k,l,b,i,z)$~\cite{Boyd-Parikh-Chu-Peleato-Eckstein-10}. By solving the
individual $s_{k,l,b,i,z}$ from~\eqref{eq:summse_admm_lag_0}, while keeping the
other variables fixed, we have
\begin{equation}
\label{eq:cohupdate_single}
    s_{k,l,b,i,z} = a_{k,l,b,i,z}^{(n)} + 
        \frac{\bar{s}_{k,l,i,z}}{|\mathcal{B}_i|} - 
        \frac{\bar{\lambda}_{k,l,i,z}^{(n)}}{|\mathcal{B}_i|}  - 
        \frac{r_{k,l,i,z}} 
             {|\mathcal{B}_i|} 
        \text{,}
\end{equation}
where $\bar{\lambda}_{k,l,i,z}^{(n)} = \sum_{g\in\mathcal{B}_i}
\lambda_{k,l,g,i,z}^{(n)}$, $a_{k,l,b,i,z}^{(n)} = \lambda_{k,l,b,i,z}^{(n)} +
\M{u}_{k,l}\herm\M{H}_{b,k}\M{m}_{b,i,z}$ and $r_{k,l,i,z} =
\sum_{g\in\mathcal{B}_i}\sqrt{w_{k,l}}\M{u}_{k,l}\herm\M{H}_{j,k}\M{m}_{g,i,z}$.
It is easy to see that~\eqref{eq:cohupdate_single}
minimizes~\eqref{eq:summse_admm_lag_0} and
satisfies~\eqref{eq:admm_sum_interference}, as derived in the following
\begin{equation}
\label{eq:admm_concensus_derivation}
    \begin{array}{rl}
        \ds 
        \sum_{b \in \mathcal{B}_i} s_{k,l,b,i,z} = 
        & \ds
        \sum_{b \in \mathcal{B}_i} a_{k,l,b,i,z}^{(n)} + \bar{s}_{k,l,i,z} - 
        \bar{\lambda}_{k,l,i,z}^{(n)}  - r_{k,l,i,z} 
        \\ = & \ds \bar{s}_{k,l,i,z}
        \text{.}
    \end{array}
\end{equation}

When we substitute each $s_{k,l,b,i,z}$ in~\eqref{eq:summse_admm_lag_0}
with~\eqref{eq:cohupdate_single}, the consensus constraints must hold as shown
in~\eqref{eq:admm_concensus_derivation}. On the other hand, the penalty
terms~\eqref{eq:admm_penalty} reduce to
\begin{equation}
    \begin{array}{rl}
    \ds \Theta_{k,l} &= 
    \ds
        \sum_{i = 1}^K\sum_{z = 1}^{L_i}\sum_{b\in\mathcal{B}_i}
            \frac{\rho}{2}|
                \frac{r_{k,l,i,z}}{|\mathcal{B}_i|} -
                \frac{\bar{s}_{k,l,i,z}}{|\mathcal{B}_i|} +
                \frac{\bar{\lambda}_{k,l,i,z}^{(n)}}{|\mathcal{B}_i|}
            |^2
    \\ \ds &= \ds
        \sum_{i = 1}^K\sum_{z = 1}^{L_i}\sum_{b\in\mathcal{B}_i}
            \frac{\rho}{2|\mathcal{B}_i|}|
                r_{k,l,i,z} - \bar{s}_{k,l,i,z} + \bar{\lambda}_{k,l,i,z}^{(n)}
            |^2
        \text{.}
    \end{array}
\end{equation}
Since, $\rho$ is an adjustable penalty constant, we can include the \ac{JP} set
sizes into it\footnote{Basically, we we would bound $\rho$ depending on the set
sizes. In any case, this is only for analytical purposes} and, thus get
\begin{equation}
    \Theta_{k,l} = 
        \sum_{i = 1}^K\sum_{z = 1}^{L_i}\sum_{b\in\mathcal{B}_i}
            \frac{\rho}{2}|
                r_{k,l,i,z} - \bar{s}_{k,l,i,z} + \bar{\lambda}_{k,l,i,z}^{(n)}
            |^2
        \text{.}
\end{equation}
Now, the same substitution for the dual update~\eqref{eq:dual_update_simple} and
having the \ac{JP} set sizes included into $\rho$, we have
\begin{equation}
\label{eq:admm_dual_update_simple_derivation}
    \begin{array}{rl}
        \ds
        \lambda_{k,l,b,i,z}^{(n+1)} &= \ds \lambda_{k,l,b,i,z}^{(n)} + 
            \sqrt{w_{k,l}}\M{u}_{k,l}\herm\M{H}_{j,k}\M{m}_{b,i,z}^{(n+1)} - 
            s_{k,l,b,i,z}^{(n+1)}
        \\ &= \ds 
            \bar{s}_{k,l,i,z} - \bar{\lambda}_{k,l,i,z}^{(n)}  - r_{k,l,i,z} 
    \text{.}
    \end{array}
\end{equation}
Since~\eqref{eq:admm_dual_update_simple_derivation} does not depend on $b$, dual
variables $\lambda_{k,l,b,i,z}\ \forall (k,l,i,z)$ are equivalent for all $b \in
\mathcal{B}_i$. Thus, we can combine all dual variables for each $(k,l,i,z)$ and
have~\eqref{eq:dual_update}.

\bibliographystyle{IEEEtran}
\bibliography{IEEEabrv,conf_short,jour_short,references}

\begin{thebibliography}{10}
\providecommand{\url}[1]{#1}
\csname url@samestyle\endcsname
\providecommand{\newblock}{\relax}
\providecommand{\bibinfo}[2]{#2}
\providecommand{\BIBentrySTDinterwordspacing}{\spaceskip=0pt\relax}
\providecommand{\BIBentryALTinterwordstretchfactor}{4}
\providecommand{\BIBentryALTinterwordspacing}{\spaceskip=\fontdimen2\font plus
\BIBentryALTinterwordstretchfactor\fontdimen3\font minus
  \fontdimen4\font\relax}
\providecommand{\BIBforeignlanguage}[2]{{%
\expandafter\ifx\csname l@#1\endcsname\relax
\typeout{** WARNING: IEEEtran.bst: No hyphenation pattern has been}%
\typeout{** loaded for the language `#1'. Using the pattern for}%
\typeout{** the default language instead.}%
\else
\language=\csname l@#1\endcsname
\fi
#2}}
\providecommand{\BIBdecl}{\relax}
\BIBdecl

\bibitem{Dahlman-Parkvall-Skold-2011}
E.~Dahlman, S.~Parkvall, and J.~Sk\"old, \emph{{4G} {LTE} / {LTE-A}dvanced for
  Mobile Broadband}.\hskip 1em plus 0.5em minus 0.4em\relax Academic Press,
  2011.

\bibitem{Shi-Razaviyayn-Luo-He-11}
Q.~Shi, M.~Razaviyayn, Z.-Q. Luo, and C.~He, ``An iteratively weighted {MMSE}
  approach to distributed sum-utility maximization for a {MIMO} interfering
  broadcast channel,'' \emph{{IEEE} Trans. Signal Processing}, vol.~59, no.~9,
  pp. 4331--4340, Sep. 2011.

\bibitem{Komulainen-Tolli-Juntti-13}
P.~Komulainen, A.~Tolli, and M.~Juntti, ``{Effective CSI Signaling and
  Decentralized Beam Coordination in TDD Multi-Cell MIMO Systems},''
  \emph{{IEEE} Trans. Signal Processing}, vol.~61, no.~9, pp. 2204--2218, 2013.

\bibitem{Tolli-Pennanen-Komulainen-TWC-11}
A.~T\"olli, H.~Pennanen, and P.~Komulainen, ``Decentralized minimum power
  multi-cell beamforming with limited backhaul signaling,'' \emph{{IEEE} Trans.
  Wireless Commun.}, vol.~10, no.~2, pp. 570--580, Feb. 2011.

\bibitem{Bogale-Vandendorpe-11}
T.~Bogale and L.~Vandendorpe, ``Weighted sum rate optimization for downlink
  multiuser {MIMO} coordinated base station systems: Centralized and
  distributed algorithms,'' \emph{{IEEE} Trans. Signal Processing}, Dec. 2011.

\bibitem{Pennanen-Tolli-Latva-Aho-11}
H.~Pennanen, A.~Tolli, and M.~Latva-aho, ``{Decentralized Coordinated Downlink
  Beamforming via Primal Decomposition},'' \emph{{IEEE} Signal Processing
  Lett.}, vol.~18, no.~11, pp. 647--650, Nov. 2011.

\bibitem{Gesbert-Hanly-Huang-Shamai-Simeone-Wei-10}
D.~Gesbert, S.~Hanly, H.~Huang, S.~Shamai~Shitz, O.~Simeone, and W.~Yu,
  ``{Multi-Cell MIMO Cooperative Networks: A New Look at Interference},''
  \emph{{IEEE} J. Select. Areas Commun.}, vol.~28, no.~9, pp. 1380--1408, 2010.

\bibitem{Zhou-Gong-Niu-WC-11}
S.~Zhou, J.~Gong, and Z.~Niu, ``{Distributed Adaptation of Quantized Feedback
  for Downlink Network MIMO Systems},'' \emph{{IEEE} Trans. Wireless Commun.},
  vol.~10, no.~1, pp. 61--67, Jan. 2011.

\bibitem{Lee-Seo-Clerckx-Hardouin-Maazarese-Nagata-Sauana-12}
D.~Lee, H.~Seo, B.~Clerckx, E.~Hardouin, D.~Mazzarese, S.~Nagata, and
  K.~Sayana, ``Coordinated multipoint transmission and reception in
  {LTE}-{A}dvanced: deployment scenarios and operational challenges,''
  \emph{{IEEE} Commun. Mag.}, vol.~50, no.~2, pp. 148--155, Feb. 2012.

\bibitem{I-Rowell-Han-Zu-Li_Pan-14}
C.~L. I, C.~Rowell, S.~Han, Z.~Xu, G.~Li, and Z.~Pan, ``{Toward green and soft:
  a 5G perspective},'' \emph{{IEEE} Commun. Mag.}, vol.~52, no.~2, pp. 66--73,
  Feb. 2014.

\bibitem{Zhang-Andrews-10}
J.~Zhang and J.~Andrews, ``Adaptive spatial intercell interference cancellation
  in multicell wireless networks,'' \emph{{IEEE} J. Select. Areas Commun.},
  vol.~28, no.~9, pp. 1455--1468, 2010.

\bibitem{Han-Yang-Wang-Zhy-Lei-13}
S.~Han, C.~Yang, G.~Wang, D.~Zhu, and M.~Lei, ``{Coordinated Multi-Point
  Transmission Strategies for TDD Systems with Non-Ideal Channel
  Reciprocity},'' \emph{{IEEE} Trans. Commun.}, vol.~61, no.~10, pp.
  4256--4270, Oct. 2013.

\bibitem{Kim-Sun-Paulraj-13}
T.~M. Kim, F.~Sun, and A.~Paulraj, ``{Low-Complexity MMSE Precoding for
  Coordinated Multipoint With Per-Antenna Power Constraint},'' \emph{{IEEE}
  Signal Processing Lett.}, vol.~20, no.~4, pp. 395--398, 2013.

\bibitem{Shi-Schubert-Boche-08}
S.~Shi, M.~Schubert, and H.~Boche, ``Rate optimization for multiuser mimo
  systems with linear processing,'' \emph{{IEEE} Trans. Signal Processing},
  vol.~56, no.~8, pp. 4020--4030, Aug. 2008.

\bibitem{Codreanu-Tolli-Juntti-Latva-aho-trsp-07}
M.~Codreanu, A.~T\"olli, M.~Juntti, and M.~Latva-aho, ``Joint design of
  {T}x-{R}x beamformers in {MIMO} downlink channel,'' \emph{{IEEE} Trans.
  Signal Processing}, vol.~55, no.~9, pp. 4639--4655, Sep. 2007.

\bibitem{Christensen-Agarwal-Carvalho-Cioffi-TWC-08}
S.~S. Christensen, R.~Agarwal, E.~Carvalho, and J.~Cioffi, ``Weighted sum-rate
  maximization using weighted {MMSE} for {MIMO-BC} beamforming design,''
  \emph{{IEEE} Trans. Wireless Commun.}, vol.~7, no.~12, pp. 4792--4799, Dec.
  2008.

\bibitem{Scutari-Facchinei-Song-Palomar-Pang-14}
G.~Scutari, F.~Facchinei, P.~Song, D.~Palomar, and J.-S. Pang, ``Decomposition
  by partial linearization: Parallel optimization of multi-agent systems,''
  \emph{{IEEE} Trans. Signal Processing}, vol.~62, no.~3, pp. 641--656, Feb.
  2014.

\bibitem{Kaleva-Tolli-Juntti-TSP16}
J.~Kaleva, A.~T\"olli, and M.~Juntti, ``Decentralized sum rate maximization
  with {QoS} constraints for interfering broadcast channel via successive
  convex approximation,'' \emph{{IEEE} Trans. Signal Processing}, vol.~64,
  no.~11, pp. 2788--2802, Jun. 2016.

\bibitem{Hong-Sun-Baligh-Luo-2013}
M.~Hong, R.~Sun, H.~Baligh, and Z.-Q. Luo, ``Joint base station clustering and
  beamformer design for partial coordinated transmission in heterogeneous
  networks,'' \emph{{IEEE} J. Select. Areas Commun.}, vol.~31, no.~2, pp.
  226--240, Feb. 2013.

\bibitem{Park-Simeone-Sahin-Shamai-13}
S.~H. Park, O.~Simeone, O.~Sahin, and S.~Shamai, ``Joint precoding and
  multivariate backhaul compression for the downlink of cloud radio access
  networks,'' \emph{{IEEE} Trans. Signal Processing}, vol.~61, no.~22, pp.
  5646--5658, Nov. 2013.

\bibitem{Dai-Yu-16}
B.~Dai and W.~Yu, ``Energy efficiency of downlink transmission strategies for
  cloud radio access networks,'' \emph{{IEEE} J. Select. Areas Commun.},
  vol.~34, no.~4, pp. 1037--1050, Apr. 2016.

\bibitem{Zhuang-Lau-2014}
F.~Zhuang and V.~Lau, ``Backhaul limited asymmetric cooperation for {MIMO}
  cellular networks via semidefinite relaxation,'' \emph{{IEEE} Trans. Signal
  Processing}, vol.~62, no.~3, pp. 684--693, Feb. 2014.

\bibitem{Lia-Hong-Liu-Luo-2014}
W.-C. Liao, M.~Hong, Y.-F. Liu, and Z.-Q. Luo, ``Base station activation and
  linear transceiver design for optimal resource management in heterogeneous
  networks,'' \emph{{IEEE} Trans. Signal Processing}, vol.~62, no.~15, pp.
  3939--3952, Aug. 2014.

\bibitem{Kaleva-Bande-Tolli-Juntti-Veeravalli-Spawc16}
J.~Kaleva, M.~Bande, A.~T\"olli, M.~Juntti, and V.~V. Veeravalli, ``{Sum Rate
  Maximizing Joint Processing with Limited Backhaul and Tree Topology
  Constraints},'' in \emph{Proc. IEEE Works. on Sign. Proc. Adv. in Wirel.
  Comms.}, Edinburgh, UK, Jul. 2016.

\bibitem{Park-Simeone-Sahin-Shamai-14}
S.~H. Park, O.~Simeone, O.~Sahin, and S.~Shamai, ``Inter-cluster design of
  precoding and fronthaul compression for cloud radio access networks,''
  \emph{{IEEE} Commun. Lett.}, vol.~3, no.~4, pp. 369--372, Aug. 2014.

\bibitem{Park-Siemone-Shain-Shamai-14_MAG}
S.~H. Park, O.~Simeone, O.~Sahin, and S.~S. Shitz, ``Fronthaul compression for
  cloud radio access networks: Signal processing advances inspired by network
  information theory,'' \emph{{IEEE} Signal Processing Mag.}, vol.~31, no.~6,
  pp. 69--79, Nov. 2014.

\bibitem{Shen-Chang-Wang-Qiu-Chi-12}
C.~Shen, T.-H. Chang, K.-Y. Wang, Z.~Qiu, and C.-Y. Chi, ``Distributed robust
  multicell coordinated beamforming with imperfect {CSI}: An {ADMM} approach,''
  \emph{{IEEE} Trans. Signal Processing}, vol.~60, no.~6, pp. 2988--3003, Jun.
  2012.

\bibitem{Kim-Shin-Sohn-Lee-12}
D.~Kim, O.-S. Shin, I.~Sohn, and K.~B. Lee, ``{Channel Feedback Optimization
  for Network MIMO Systems},'' \emph{{IEEE} Trans. Veh. Technol.}, vol.~61,
  no.~7, pp. 3315--3321, Sep. 2012.

\bibitem{Jaramillo-Ramirez-Kountouris-Hardouin-15}
D.~Jaramillo-Ramírez, M.~Kountouris, and E.~Hardouin, ``Coordinated
  multi-point transmission with imperfect {CSI} and other-cell interference,''
  \emph{{IEEE} Trans. Wireless Commun.}, vol.~14, no.~4, pp. 1882--1896, Apr.
  2015.

\bibitem{Jose-Ashikmin-Marzetta-Vishwanath-WC-11}
J.~Jose, A.~Ashikhmin, T.~L. Marzetta, and S.~Vishwanath, ``Pilot contamination
  and precoding in multi-cell tdd systems,'' vol.~10, no.~8, pp. 2640--2651,
  Aug. 2011.

\bibitem{Shi-Berry-Honig-TSP-14}
C.~Shi, R.~Berry, and M.~Honig, ``Bi-directional training for adaptive
  beamforming and power control in interference networks,'' \emph{{IEEE} Trans.
  Signal Processing}, vol.~62, no.~3, pp. 607--618, Feb. 2014.

\bibitem{Xu-Guo-Honig-TSP-15}
M.~Xu, D.~Guo, and M.~L. Honig, ``Distributed bi-directional training of
  nonlinear precoders and receivers in cellular networks,'' \emph{{IEEE} Trans.
  Signal Processing}, vol.~63, no.~21, pp. 5597--5608, Nov. 2015.

\bibitem{Kaleva-Berry-Honig-Tolli-Juntti-ICASSP-14}
J.~Kaleva, R.~Berry, M.~Honig, A.~Tolli, and M.~Juntti, ``{Decentralized sum
  MSE minimization for coordinated multi-point transmission},'' in \emph{Proc.
  IEEE Int. Conf. Acoust., Speech, Signal Processing}, May 2014, pp. 469--473.

\bibitem{Luo-Zhang-08}
Z.~Luo and S.~Zhang, ``Dynamic spectrum management: Complexity and duality,''
  \emph{{IEEE} J. Select. Areas Commun.}, vol.~2, no.~1, pp. 57--73, Feb. 2008.

\bibitem{Boyd-Vandenberghe-04}
S.~Boyd and L.~Vandenberghe, \emph{Convex Optimization}.\hskip 1em plus 0.5em
  minus 0.4em\relax Cambridge, UK: Cambridge University Press, 2004.

\bibitem{Boyd-Parikh-Chu-Peleato-Eckstein-10}
S.~Boyd, N.~Parikh, E.~Chu, B.~Peleato, and J.~Eckstein, ``Distributed
  optimization and statistical learning via the alternating direction method of
  multipliers,'' \emph{Foundations and Trends in Machine Learning}, vol.~3,
  no.~1, pp. 1--122, 2010.

\bibitem{Boyd-EE364b-PrimDualDecomp-07}
S.~Boyd, ``Primal and dual decomposition,'' 2007, [Online]. Available:
  \url{http://www.stanford.edu/class/ee364b/lectures/decomposition_slides.pdf}.

\bibitem{Luo-Davidson-Giannakis-Wong-04}
Z.~Q. Luo, T.~N. Davidson, G.~Giannakis, and K.~M. Wong, ``Transceiver
  optimization for block-based multiple access through {ISI} channels,''
  \emph{{IEEE} Trans. Signal Processing}, vol.~52, no.~4, pp. 1037--1052, Apr.
  2004.

\bibitem{Hong-Luo-Razaviyayn-14}
M.~{Hong}, Z.-Q. {Luo}, and M.~{Razaviyayn}, ``{Convergence Analysis of
  Alternating Direction Method of Multipliers for a Family of Nonconvex
  Problems},'' \emph{ArXiv e-prints}, Oct. 2014.

\bibitem{Haykin-96}
S.~Haykin, \emph{Adaptive Filter Theory}, 3rd~ed.\hskip 1em plus 0.5em minus
  0.4em\relax Upper Saddle River, NJ, USA: Prentice Hall, 1996.

\end{thebibliography}

\end{document}